\documentclass{article}
\usepackage{codefuse_tech_report}
\usepackage[colorlinks = true,
            linkcolor = blue,
            urlcolor  = blue,
            citecolor = blue,
            anchorcolor = blue]{hyperref}
\usepackage{microtype}
\usepackage{xurl}
\usepackage{booktabs}
\usepackage{enumitem}
\usepackage{multicol}
\usepackage{multirow}
\usepackage{graphicx}
\usepackage{authblk}
\usepackage{adjustbox}

\usepackage{array}
\usepackage{amsmath}
\usepackage{amssymb}
\usepackage{amsthm}
\usepackage{mathtools}

\usepackage{caption}
\usepackage{stfloats}
\usepackage{makecell}
\usepackage{threeparttable}
\usepackage{longtable}
\usepackage{colortbl}
\usepackage{pifont}
\usepackage{fontawesome5}

\usepackage{xcolor}
\usepackage{tcolorbox}
\tcbuselibrary{skins,breakable}

\usepackage{xspace}
\usepackage{listings}
\usepackage[normalem]{ulem}     % \uline used by \myreason

\usepackage{tikz}

\usepackage[capitalize,noabbrev]{cleveref}

\newcommand*\circled[1]{\tikz[baseline=(char.base)]{
        \node[shape=circle,draw,inner sep=0.5pt] (char) {\small#1};}}

\newcommand{\proj}{\textsc{DeBench}\xspace}

\newcommand{\myparagraph}[1]{%
  \par\noindent\textbf{\textit{#1.}}\quad\ignorespaces}
\newcommand{\myparagraphnew}[1]{%
  \par\noindent\textbf{\textit{#1?}}\quad\ignorespaces}

\providecommand{\eg}{\textit{e.g.}\xspace}

\newcommand{\pz}[1]{}
\newcommand{\yh}[1]{}

\newcounter{findingnumber}
\newcommand{\finding}[1]{%
\begin{tcolorbox}[colback=gray!8,colframe=gray!60,boxsep=2pt,left=4pt,right=4pt,top=2pt,bottom=2pt,arc=2pt]
\textbf{Finding \refstepcounter{findingnumber}\thefindingnumber}: #1
\end{tcolorbox}}

\Crefname{section}{Section}{Sections}
\crefname{section}{Section}{Sections}
\Crefname{subsection}{Section}{Sections}
\crefname{subsection}{Section}{Sections}
\Crefname{table}{Table}{Tables}
\crefname{table}{Table}{Tables}
\Crefname{figure}{Figure}{Figures}
\crefname{figure}{Figure}{Figures}

\title{CODEFUSE-\proj: An Empirical Study on  Readability, Recompilability, and Functionality}

\author{%
Puzhuo Liu$^{1,2}$
~~Yuhan Huang$^{1,3}$
~~Jianlei Chi$^{3}$
~~Peng Di$^{1}$
~~Yu Jiang$^{2}$
\\
$^1$Ant Group\ \ \ $^2$Tsinghua University\ \ \ $^3$Xidian University\ \ \\
}

\colmfinalcopy

\begin{document}

\maketitle

\begin{abstract}
Binary decompilation is a core technology in software security and reverse engineering, aiming to recover executable files into high-level source code. Existing evaluations rely on syntactic similarity metrics (lines of code, character-level matching) or single-axis readability scores, which cannot measure the practical engineering value of decompiled code. To address this, we propose a reusability-driven evaluation paradigm that systematically assesses decompiler capabilities along three orthogonal dimensions---readability, recompilability, and functionality---and ask not just \emph{whether} but \emph{where} decompiled code is reusable.

We build \proj, the first automated framework for multidimensional decompilation evaluation. \proj contains 240 atomic test functions grouped into 8 dimensional source files and compiled into 640 binaries, and integrates LLM-as-judge readability scoring (URAF, 18 sub-dimensions), iterative compile-and-repair under a fixed 50-iteration budget, and Frida-based differential dynamic tracing at program / function / instruction granularities. Through an empirical study of five mainstream decompilers and three independent repair LLMs, we report four contributive findings that are not visible to single-axis evaluations: (1) the \emph{reusability cliff} is steep and reproducible---the best decompiler--LLM pair reaches 22.3\% Exact+Partial program-level behavioral overlap but only 1.2\% exact stdout match, a $\sim$50-percentage-point cliff below recompilability; (2) compilation choices that maximize readability are \emph{not} the ones that maximize functionality---aggressive \texttt{O3} yields the lowest readability but the highest functionality, and Clang yields lower readability than GCC but $2.6\times$ higher functionality; (3) the decompiler-versus-LLM gap is asymmetric---the cross-decompiler swing at functional level ($20\times$) dwarfs the cross-LLM swing ($1.6\times$), so engineering investment should target decompiler engines, not larger repair models; (4) failures decompose into three categories with distinct reparability---syntactic noise (repairable), type-system collapse ($\sim$19\% of all repair errors and the dominant non-syntactic cost), and irreversible upstream losses (ARM64 relocation idioms and C++ ABI features) that no repair LLM can compensate for.

Our study reframes decompiler quality as a multi-axis continuum rather than a single-axis score, and points to three priority directions for decompiler-side engineering: composite-type reconstruction that survives symbol stripping, IR-level symbolic relocation, and C++ ABI-aware front-end passes.
\end{abstract}

\section{Introduction}
Binary program\footnote{
This paper focuses on decompiling C-language executables. 
For brevity, 
binary programs in this paper refer to C-language executable programs.} 
analysis has become indispensable in modern software practice. 
First, the source code of many targets, \eg, commercial software,
 firmware, and malware, is unavailable, 
 making third-party binary-based analysis and testing the default 
 choice~\cite{liu2022finding,liu2023fits,liu2025llm,wu2024your,bridge2026}.
Second, even when source code exists, today’s compiler toolchains and build systems are highly complex and heterogeneous, creating serious compatibility challenges for source-level analyzers~\cite{aidasso2025build,mcintosh2011empirical,zhou2024plankton,liu2026bit}. 
Moreover, source code has not yet undergone compiler transformations and optimizations, so source-level conclusions do not always reflect the behavior of the code that actually runs~\cite{zhong2025understanding,xu2023silent,sun2016toward}. 
Instead, binary analysis operates directly on the machine-executed artifact, avoiding many compatibility issues and improving both behavioral consistency and the verifiability of analysis results~\cite{meng2016binary,eschweiler2016discovre,richter2023train}.

Decompilation lies at the heart of many binary analysis tasks: it reverses low-level machine code into a high-level, human-readable representation~\cite{cifuentes1995decompilation,tan2024llm4decompile,meng2016binary}.
Traditionally, tools such as IDA Pro~\cite{ida}, Ghidra~\cite{ghidra}, and RetDec~\cite{retdec} have focused on producing readable “pseudocode” that approximates the original program’s semantics and structure, primarily to assist human reverse engineers in auditing binaries. 
With the advent of large language models (LLMs) 
and the growing demand for advanced binary rewriting, 
expectations have risen substantially~\cite{hu2024degpt,wong2025decllm,tan2024llm4decompile,liu2025function}. 
The community now aims not only for readable output,
but for syntactically correct, recompilable decompiled code. The ultimate goal is to generate a new binary that is functionally equivalent to the original. Meeting this stronger standard of reusability would unlock a wide range of downstream applications traditionally limited to source code, including automated patching, cross-architecture migration, and the use of powerful static analysis frameworks.

Despite this demand, objectively, reliably, and comprehensively evaluating the capabilities of modern decompilers remains a gap.
We observed (see \cref{subsec:moti}) that the decompiled code often presents a visually different representation, which directly impacts readability, but its impact on recompilability and functionality requires further investigation.
Furthermore, there are clear differences between decompiled code and source code (due to information loss caused by compiler transformations and optimizations).
However, which information gaps truly affect the readability, compatibility, and functionality of decompilers has not yet been systematically revealed.
This has caused great confusion for normal users, who want to use decompiled code for various downstream applications, 
and is a research gap that is not fully disclosed by decompiler developers.

This research aims to systematically determine the true capability of modern binary decompilers.
This work seeks to answer the following key research question: “\textit{To what extent can modern decompilers produce readable, recompilable, and functionally equivalent code?}” 
To answer this, we establish a reusability-driven evaluation paradigm that prioritizes the practical utility of decompiled output over superficial similarity. 
Our approach addresses three limitations in existing evaluation methodologies.
\begin{itemize}[leftmargin=10pt,topsep=0pt,itemsep=0pt,parsep=0pt]
    \item In response to hand-crafted samples with simple logic and low fidelity, as well as the risks of data leakage and licensing issues in real program sets and the high cost and poor targeting of large-scale collection~\cite{10.1145/3650212.3652144,tan2025decompilebench}, 
we construct a blended test suite grounded in LLM knowledge, expert experience, and real-world programs. 
It provides integration evaluations that cover key reverse engineering challenges—complex control flow, indirect branching, pointers and composite types, calling conventions, and cross-module interactions—while decomposing them into fine-grained unit cases. We compile across multiple compilers, optimizations, and architectures to improve diversity and fidelity.
\item To correct the misalignment of traditional metrics that emphasize superficial consistency (e.g., code/variable counts, structural matching) while overlooking structural changes introduced by optimization and deviating from practical needs~\cite{10.1145/3520312.3534867,10.1145/3650212.3652144}, 
we align evaluation targets with real reuse scenarios: readability and naming quality are scored by LLMs using the original source as reference; syntactic correctness and recompilability are quantified by recompilation and weighted by the number and types of diagnostics; functionality is assessed via instrumentation and dynamic execution by comparing contextual differences in standard library calls and side effects. 
\item To overcome dataset-driven evaluations that are reproducible yet ineffective, subjective and unscalable human reviews, and semi-automated (LLM-assisted) reviews lacking checks for grammar and equivalence~\cite{10.1145/3650212.3652144,tan2025decompilebench}, we implement an end-to-end automated pipeline that integrates LLMs, compiler toolchains, and instrumentation frameworks. 
This pipeline minimizes human intervention, reduces subjectivity, improves scalability and reproducibility, and supports batch evaluation of decompilers.
\end{itemize}

To address these gaps, we present \proj, a reusability-driven evaluation framework that systematically assesses binary decompilers across readability, recompilability, and functionality. Using a 240-function atomic benchmark grouped into 8 dimensional source files and compiled into $8\!\times\!2\!\times\!5\!\times\!2\!\times\!4=640$ distinct binaries across compilers, optimization options, debug-symbol settings, and target architectures, we evaluate five mainstream whole-program decompilers (IDA, Ghidra, RetDec, Angr, BinaryAI) through a fully automated pipeline; the readability and function-level functionality components additionally extend to LLM-based decompilers, whose function-granularity output is not directly comparable on program-level recompilability.
Our contributions are as follows:
\begin{itemize}[leftmargin=10pt,topsep=0pt,itemsep=0pt,parsep=0pt]
    \item Reusability-driven evaluation paradigm.
We propose a reusability-centered evaluation paradigm for decompilers that goes beyond syntactic similarity. It quantitatively assesses decompiled code along three orthogonal engineering dimensions: readability, recompilability, and functionality, providing a more rigorous basis for judging its practical utility.

\item End-to-end automated infrastructure.
We design and implement \proj, the first automated framework for comprehensive decompilation evaluation, featuring (i) a benchmark suite of 240 atomic-cases across 8 code dimensions and diverse compilation settings (640 binaries), and (ii) an integrated pipeline combining LLM-as-judge semantic scoring, LLM-based iterative compile-and-repair, and Frida-based dynamic behavior verification.

\item Empirical insights and practical guidelines.
We perform an empirical study of five mainstream decompilers, characterizing their capability limits and failure modes. The framework is designed for three concrete user groups: \textit{decompiler developers}, who can use the per-dimension and per-failure-mode breakdowns to prioritize engineering work (e.g., type-system reconstruction, ARM64 relocation handling, C++ ABI recovery); \textit{security analysts}, who can use the per-decompiler readability and functionality profiles to pick the right tool for a given target; and \textit{compiler/program-analysis researchers}, who can use the open artifact and the failure-mode taxonomy as ground for studying information loss across the lowering--raising boundary. Our results provide scenario-based tool selection guidance and identify core bottlenecks, such as semantic opacity and type-system degradation, that define the path from syntactic translation to semantic reconstruction.
\end{itemize}

\section{Preliminaries}
\begin{figure}[!t]
	\centering
	\includegraphics[width=\columnwidth]{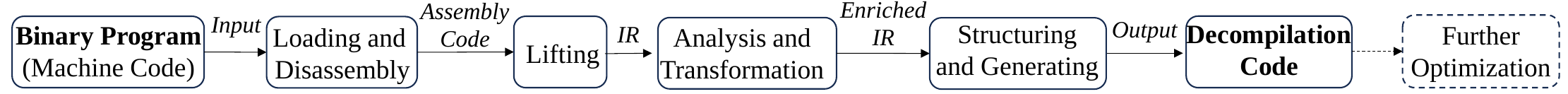}
	\caption{The workflow of modern binary decompiler.}
	\label{fig_decompie}
\end{figure}

Decompilation converts compiled machine code back into a high-level source representation via a common workflow (\cref{fig_decompie}),  though implementations differ.

\myparagraph{Loading and Disassembly}
The \textit{loader} parses the executable format, resolves sections, metadata, and handles packed or position-independent code. The \textit{disassembler}  then turns bytes into assembly, discovers code and function boundaries, and deals with indirect jumps and calls using heuristics and data-flow reasoning.

% A \textit{loader} parses the executable format (e.g., ELF, PE), resolves sections and metadata (symbols, relocations, imports/exports), and accounts for instruction set architecture (ISA), application binary interface (ABI), endianness, and position-independent code; packed or encrypted binaries may require unpacking or tracing. 
% The \textit{disassembler} then converts bytes to assembly and discovers code regions and function boundaries using recursive descent or linear sweep with heuristics for prologues, jump tables, and exception frames. Indirect transfers (function pointers, virtual calls, computed jumps) are addressed with pattern matching, data-flow reasoning, and occasionally dynamic profiling to avoid code–data confusion.

\myparagraph{Lifting to an Intermediate Representation (IR)}
Architecture-specific instructions are translated into a uniform IR such as Pcode~\cite{ghidra}. The IR makes registers, memory, and side effects explicit, organizes code into basic blocks and a control-flow graph (CFG), and is often converted to static single assignment form for easier analysis.

% Architecture-specific instructions are translated into a uniform IR (e.g., Microcode~\cite{ida}, Pcode~\cite{ghidra}) that makes registers, memory, flags, and side effects explicit, organizes code into basic blocks, and builds a control-flow graph (CFG). 
% Precise modeling of calling conventions, stack discipline, and ABI constructs is required; 
% many systems further convert IR to static single assignment (SSA) to simplify data-flow analysis.

% Instruction sequences specific to particular CPU architectures (e.g., x86, ARM) are mapped into an abstract form composed of simpler, generic operations, thereby decoupling the core analysis engine from the source architecture and substantially simplifying subsequent analyses. The IR uniformly captures operations, control flow, and side effects, typically making registers, memory accesses, and flags explicit, and organizes straight-line code fragments into basic blocks that terminate with control-flow instructions, which are then assembled into a control-flow graph (CFG). The lifting process further requires precise modeling of calling conventions, stack discipline, and ABI-specific constructs (such as thread-local storage and variadic arguments) to ensure semantic fidelity and analyzability; to facilitate data-flow analysis and optimization, many systems additionally convert the IR into Static Single Assignment (SSA) form.

\myparagraph{Analysis and Transformation} 
Various analyses recover high-level structure and semantics: control-flow analysis refines CFGs and reconstructs loops, switches, and exception edges; data-flow and pointer analyses clarify values and memory relationships; inter-procedural analysis resolves indirect calls; type and variable recovery infers data types, variables, and object layouts; function recovery identifies boundaries, parameters, and return values.

\myparagraph{Structuring and Generating}
The enriched IR is restructured into high-level constructs (if/else, loops, switches), with gotos used when necessary, and then pretty-printed—typically as C—using the inferred types, variables, and signatures.

% The final phase of decompilation involves structuring the enriched IR and generating the target high-level source code. 
% The IR is structured into high-level constructs (if/else, loops, switches) using region-based analysis and dominance relations, with gotos as a fallback for irreducible graphs. 
% The final code generator pretty-prints the structured IR (using inferred types, variables, and signatures), most commonly in C.

% The final code \textit{generation} step translates the structured IR into a human-readable high-level language, most commonly C. 
% This step is akin to "pretty-printing" the recovered program semantics, utilizing the identified control structures, named variables with inferred types, and function signatures to produce a syntactically correct and coherent source code file.

\myparagraph{Further Optimization} Decompilers may refine output by using debug information, recognizing common patterns (e.g., standard library calls), combining static and dynamic analysis, and improving type and name recovery, making the result more readable and closer to the original source.

\section{Motivating Example}\label{subsec:moti}

A decompiled program may appear structurally clean and easy to understand, yet still deviate from the original semantics in subtle but critical ways. Conversely, another output may look awkward and difficult to read, while preserving the source semantics much more faithfully. This mismatch motivates the need for a benchmark that evaluates decompilation quality from multiple complementary dimensions rather than relying on readability alone.

\begin{figure}[!h]
    \centering
    \includegraphics[width=\linewidth]{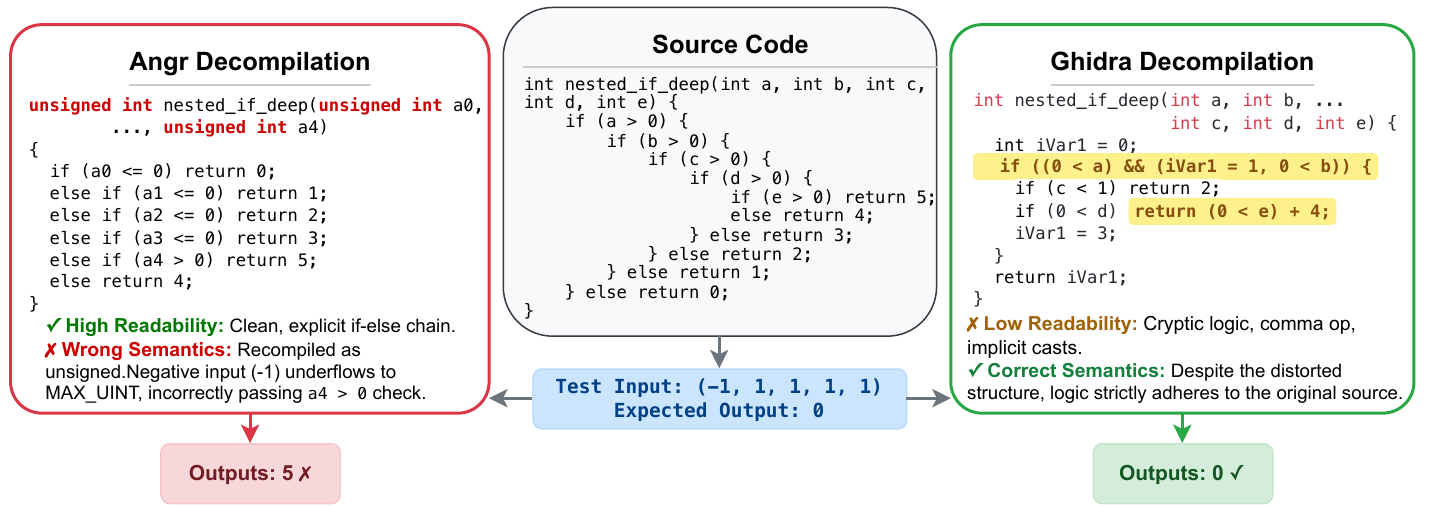}
    \caption{The decompiled result of the same function.
    Angr appears cleaner but returns an incorrect value; Ghidra, despite being harder to read due to expression complexity, preserves correct semantics and returns the expected value. 
   }
    \label{fig:motivation}
\end{figure}

\cref{fig:motivation} shows that the function (source code) \texttt{nested\_if\_deep} implements a five-level nested conditional over five signed integer inputs. Its behavior is straightforward: it checks whether $a$, $b$, $c$, $d$, and $e$ are greater than zero in sequence, and returns the index of the first variable that fails the positivity test. Because the original source uses signed integers throughout, negative values are semantically important and directly affect the control flow.

 % Case~A (Angr) looks cleaner and returns~5, because all parameters are incorrectly recovered as \mycode{unsigned\,int}.
 %    Case~B (Ghidra) is harder to read because of comma-expression side effects and boolean arithmetic, yet still returns the expected value~0 and remains closer to the source semantics.
 %    Highlighted lines mark the critical constructs behind this mismatch between appearance and semantic quality.

\myparagraph{Angr --- Readable yet Semantically Incorrect}
The decompiled output is visually clean. It recovers a compact and highly regular if-else if chain, with explicit branch structure and little syntactic noise. From a human reader's perspective, this code is easier to follow than the original deeply nested version and would likely be judged as having high readability. However, a closer inspection reveals a critical semantic error: all parameters are recovered as \texttt{unsigned int} rather than \texttt{int}. As a result, the intended signed comparisons are silently transformed into unsigned ones. For input domains involving negative values, this changes the program behavior fundamentally. In other words, although the code looks reasonable, its logic is no longer equivalent to the source.

\myparagraph{Ghidra --- Unreadable yet Semantically Faithful}
The decompiled output presents the opposite case. Its structure is clearly less readable: it introduces an auxiliary variable, uses a comma expression with side effects inside a condition, and encodes one branch as a boolean arithmetic expression \texttt{return (0 < e) + 4}. These constructs increase cognitive burden and make the control flow harder to understand at a glance. Nevertheless, after aligning each branch with the source logic, we find that this decompiled version preserves the original semantics correctly, including cases involving negative inputs. That is, despite its poor surface form, its underlying behavior remains faithful to the source program.

\myparagraph{Implication}
This example highlights a key challenge in binary decompilation evaluation: readability and semantic fidelity can diverge substantially. A decompiler output that appears polished may still be semantically wrong, while a seemingly ugly output may actually be behaviorally correct. Moreover, even if a decompiled program can be repaired and recompiled, recompilability alone still does not guarantee semantic correctness. Therefore, evaluating decompilers along a single axis is insufficient.

This observation directly motivates the design of \proj. Instead of conflating different notions of quality, \proj decomposes decompilation assessment into three complementary dimensions:
\textbf{Readability}, which captures how understandable the recovered code is to humans;
\textbf{Recompilability}, which measures whether the decompiled output can be repaired and successfully recompiled;
\textbf{Functionality}, which validates whether the repaired output preserves the original runtime behavior.
By jointly considering these dimensions, \proj provides a more faithful and fine-grained evaluation of decompiler quality.

\myparagraph{Worked end-to-end output for \texttt{nested\_if\_deep}}
To make the three pipelines concrete on a single case, we briefly report what \proj produces for the function in \cref{fig:motivation}.
For \textbf{Angr}: URAF assigns a relatively high \emph{Structural Intelligibility} score (the if-else chain is clean) but a low \emph{Type-System Fidelity} score (parameters recovered as \texttt{unsigned int} when the source uses \texttt{int}); the recompilability pipeline reaches \textit{Full Success} in only a few iterations (the syntax is valid); the functionality pipeline classifies it as \textit{Fail} at the program level, because the case driver feeds negative inputs and the recompiled binary returns a different value than the original.
For \textbf{Ghidra}: URAF assigns a lower \emph{Structural Intelligibility} (comma-expression side effects and boolean arithmetic increase cognitive load) but a higher \emph{Type-System Fidelity} (signed types preserved); the recompilability pipeline also reaches FS; the functionality pipeline classifies it as \textit{Exact Stdout}, because the recompiled binary matches the original on all driver inputs including the negative-input branches.
The Angr--Ghidra split on this single case is exactly the divergence URAF was designed to expose: high \textit{Structural Intelligibility} with low \textit{Type-System Fidelity} predicts a tool that produces clean-looking code that is nonetheless behaviorally wrong, which is precisely what the functionality pipeline confirms downstream.

\section{Study Design}
We present a framework that assesses binary decompilers along three orthogonal perspectives. To avoid the term-conflation common in earlier evaluation literature, we fix the meaning of each term used in the rest of the paper:

\begin{itemize}[leftmargin=10pt,topsep=2pt,itemsep=1pt,parsep=0pt]
  \item \textbf{Readability} is the degree to which a human reverse engineer can comprehend the decompiled C, scored \emph{relative} to the corresponding source. We measure it via URAF, an LLM-judged 18-sub-dimension rubric organized into 5 levels. ``Readability'' in this paper never means a static metric (LOC, cyclomatic complexity); it means the URAF score.
  \item \textbf{Recompilability} is the ability to obtain a valid linkable binary from decompiler output through bounded LLM-driven repair. The metric is the three-tier outcome (FS/LF/CF) under a 50-iteration repair budget, complemented by the normalized fraction of initial errors removed. ``Recompilability'' never means ``compiles in one shot.''
  \item \textbf{Functionality} is the runtime behavioral equivalence between the recompiled binary and the original under the same case driver, observed at three granularities (program / function / instruction). The headline metric is the program-level four-category outcome (Exact / Partial / Fail / Unsupported). ``Functionality'' never means ``passes all unit tests in some external suite''---we use an \emph{in-suite} driver designed to exercise the function under test.
  \item \textbf{Ground truth} in this paper is the \emph{original source code} and the \emph{original binary's runtime behavior}: source for readability and recompilability comparison, runtime traces for functionality. The LLM judge does \emph{not} act as ground truth; it acts as a rater whose judgement is constrained by the source-code reference and cross-checked against the deterministic recompilability and functionality oracles.
\end{itemize}

\begin{figure}[!t]
    \centering

    \includegraphics[width=\linewidth]{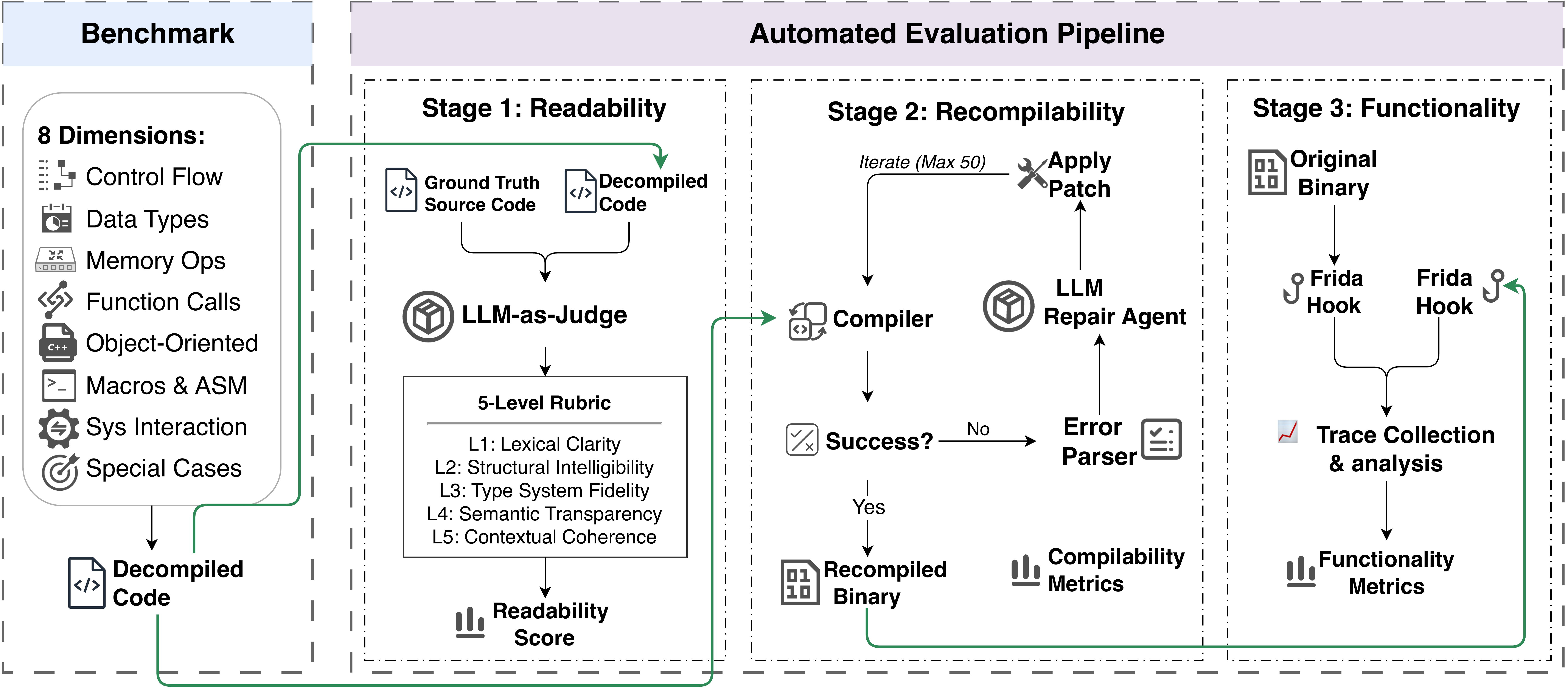}
    \caption{The evaluation pipeline of \proj.}
    \label{fig:pipeline}
\end{figure}

\cref{fig:pipeline} illustrates our integrated process: benchmark design followed by the three-stage evaluation.

\subsection{Benchmark Design}\label{subsecbench}
Existing benchmarks are plagued by entangled logic, coarse-grained annotations, and domain bias. To overcome these limitations, we present \proj, a fine-grained, function-level suite for systematic binary decompiler evaluation. Our design adheres to Atomicity (single-primitive test cases) and Progressive Complexity (scalable difficulty from trivial to pathological),

% and Scenario (authentic coding patterns across diverse domains).

\myparagraph{Atomicity}
To ensure precise localization of specific language features that trigger decompiler failures, test cases in \proj are designed with strict atomicity. 
This design effectively mitigates the challenges of tracing error propagation within complex logic. 
Specifically, \proj comprises 240 cases implemented as independent functions, meticulously curated from C/C++ language specifications and real-world reverse engineering patterns. 
Unlike wild samples, which are often difficult to diagnose, \proj adopts a bottom-up construction methodology. 
As shown in \cref{tab:bench}, each case is a function with input and output to facilitate subsequent execution monitoring. 
Furthermore, the cases are categorized along 8 different dimensions, resulting in 8 independent test files for dimensional evaluation.
To guarantee execution isolation, test files are compiled into independent executable binaries free of unresolved external symbols, thereby eliminating link-time dependencies. 
Furthermore, they avoid implicit global state dependencies that could introduce non-deterministic execution paths. 
Finally, each function is designed to produce deterministic output or return values, facilitating verification by automated testing tools.

\begin{table}[!h]
\centering
\caption{Test cases composition across eight dimensions.}
\label{tab:bench}
\resizebox{\columnwidth}{!}{%
\begin{tabular}{lll}
\hline
Dimension                & Cases & Focus Area                                                                \\ \hline
Control Flow             & 45    & Branching, irreducible loops, recursion, and computed jumps               \\
Data Types \& Variables  & 48    & Primitive types, arrays, pointer chains, and composite layouts            \\
Memory Operations        & 44    & Stack layout, heap patterns, and aliasing                                 \\
Function Calls           & 36    & Calling conventions, callbacks, parameter passing, and return semantics   \\
Object-Oriented C++      & 12    & Class lifecycles, inheritance, virtual dispatch, and templates            \\
Compile-Time Specialization & 17    & Constant/idiom recovery from macro-expanded code, surviving-branch reconstruction, inline assembly, and SIMD intrinsics \\
System Interaction       & 25    & Standard libraries, system calls, IPC, and threading                      \\
Special Challenges       & 13    & Opaque predicates, optimization-induced patterns, and edge cases          \\ \hline
\end{tabular}%
}
\end{table}

A note on the \textit{Compile-Time Specialization} dimension is in order. Macro expansion and conditional compilation are resolved by the preprocessor, so the corresponding binary is identical to one produced from already-expanded source. We therefore do not test these as new lowering challenges; instead, we test whether decompilers can \emph{re-introduce} source-level structure that the preprocessor erased: recovering magic constants as named macros (e.g., recognizing \texttt{1024} as \texttt{MAX\_SIZE} via signature databases or comments), inferring that a surviving branch came from an \texttt{\#ifdef}, and recovering inline-assembly/SIMD intrinsics that bypass the regular IR lifting path. These are real decompiler-side capabilities, distinct from raw control-flow recovery, and modern tools such as IDA and Ghidra implement parts of them.

\myparagraph{Progressive Complexity}
We categorize the test cases into five difficulty levels (\textit{L1}–\textit{L5}) to clearly distinguish decompilers that can handle conventional code from those capable of dealing with complex code.
Traditional metrics such as lines of code or cyclomatic complexity are insufficient as implementation-agnostic, generalizable measures of decompilation difficulty: a very short function that heavily relies on undefined behavior can be extremely hard to decompile correctly, whereas a very long but strictly sequential function may be trivial.
To approximate difficulty in a structured way, we adopt a three-view scoring model. Each test case is rated along three orthogonal views, and the final difficulty level is determined by their weighted sum. The weights below reflect our domain judgment informed by common decompiler failure patterns observed during benchmark development.
\begin{itemize}
[leftmargin=10pt,topsep=0pt,itemsep=0pt,parsep=0pt]
    \item Control/Data-Flow Complexity. This is the dominant factor, as the core task of decompilation is to reconstruct the control-flow graph and infer types and data flow from unstructured assembly. We assign it the highest weight because structural complexity (e.g., abundant goto statements, complex pointer aliasing, deeply nested branches/loops) poses the most fundamental algorithmic challenges to static analysis engines.
    \item Optimization Resistance. Compiler optimizations (e.g., -O3) often eliminate or rewrite high-level syntactic structures. If a feature is easily optimized away (such as a simple constant-foldable loop), it poses little challenge in the resulting binary. This view measures the persistence of a feature in the compiled binary. 
    %We assign a 30% weight because it determines whether the challenge actually manifests in the input.
    \item Semantic Loss Risk. This view captures the inherent ambiguity of code patterns. Certain constructs (e.g., null pointer casts, union accesses) are mathematically non-invertible or highly ambiguous without additional context, imposing an intrinsic upper bound on the correctness of decompilation. 
    %We weight this at 30% to characterize the theoretical upper limit of recoverable semantics.

\end{itemize}

\subsection{Evaluation Pipeline}
\begin{table*}[!th]
\centering
\caption{Readability Evaluation Levels}
\label{tab:readiablity}
\resizebox{\textwidth}{!}{%
\begin{tabular}{ll}
\hline
 \textbf{Level}                  & \textbf{Rationale}                                                                                                                \\ \hline
 \textbf{Lexical Clarity}                  & Identifier recovery (e.g., packet\_size vs. v1) minimizes cognitive mapping burden; foundational for all subsequent comprehension \\
\textbf{Structural Intelligibility}            & Natural control-flow reconstruction; excessive goto statements or irreducible loops disrupt mental models                         \\
\textbf{Type-System Fidelity}                 & Accurate type recovery (int* vs. char*, struct layouts) underpins correct static analysis and vulnerability identification        \\
 \textbf{Semantic Transparency}                & Recognition of algorithmic intent (e.g., CRC32, linked-list traversal) and documentation of undefined behavior                    \\
 \textbf{Contextual Coherence}               & Integration with global context: cross-references, module organization, interface consistency                                     \\ \hline
\end{tabular}%
}
\end{table*}
\subsubsection{Readability}
Quantifying the readability of decompiled code is inherently subjective. 
Prior work has relied on manual user studies—expensive, unscalable, and prone to bias—or on coarse static metrics, which do not reflect the actual cognitive effort required for a human analyst (see also the difficulty discussion above). To close this gap, we introduce the \textit{Unified Readability Assessment Framework (URAF)}, a systematic framework that evaluates decompiler output directly against the ground-truth source. URAF is grounded in how reverse engineers actually read code. We assume a hierarchical comprehension process: analysts first parse lexical tokens (identifiers), then reconstruct control-flow and structure, and finally reason about high-level semantics. 
To scale this evaluation, we adopt an LLM-as-a-Judge approach to reason about clarity and intent. 
We use LLMs driven by a strict, rubric-based system prompt that forces the model to act as an objective auditor. 

The evaluation is comparative: instead of asking whether the decompiler outputs perfect code in absolute terms, we ask to what extent it recovers the semantic intent and readable structure of the original source code; the judge receives both the decompiled code and the source as input. The five levels in \cref{tab:readiablity} are decomposed into a total of \textbf{18 sub-dimensions} that the judge scores independently: \textit{Lexical Clarity} covers variable naming, function naming, literal clarity, and noise reduction; \textit{Structural Intelligibility} covers control-flow naturalness, function decomposition, code linearization, and redundancy elimination; \textit{Type-System Fidelity} covers type precision, composite-type recovery, and type safety; \textit{Semantic Transparency} covers decompiler-emitted comment quality, idiom recognition, anomaly indication, and assertion/contract recovery; \textit{Contextual Coherence} covers cross-reference clarity, module organization, and build compatibility. The complete rubric---scoring criteria, focus questions, and worked examples for every sub-dimension---is shipped with the artifact and is too long to fit here.

We use a 1--10 scale rather than a coarser 1--5 scale because each tier is anchored to a five-band qualitative description (10 ``as readable as the original'', 8--9 ``minor issues'', 6--7 ``moderate issues'', 4--5 ``significant issues'', 2--3 ``severe issues'', 1 ``unusable''); the finer scale lets the judge break ties within a tier without forcing it onto a coarser bucket and is consistent with prior LLM-as-judge work. Each level score is the arithmetic mean of its sub-dimension scores; the overall score is the arithmetic mean of the five level scores. We deliberately use equal weights at the level of aggregation: pilot runs with the prompt's nominal weights (Lexical 30\%, Structural 25\%, Type 20\%, Semantic 15\%, Contextual 10\%) reordered no decompiler pair, so the equal-weight choice is reported for transparency rather than tuned for a specific outcome. This structure isolates strengths and weaknesses---for example, a tool may score well on \textit{Structural Intelligibility} but poorly on \textit{Type-System Fidelity}---a nuance a single aggregate metric would obscure.

To mitigate single-judge bias, we run \emph{three} LLM judges (GLM-4.7, Qwen3.5-Plus, MiniMax-M2.5) independently on every (decompiler, binary) pair and report all three. The judges are treated as independent raters: for the readability tables we report the per-judge score in addition to the cross-judge mean, and we use cross-judge agreement (rank-stability of the decompiler ordering and the spread of absolute scores) as our inter-rater check. Where they disagree on absolute level we note it; where they agree on rank order we treat the conclusion as robust to judge identity.

\myparagraph{Triangulation against Deterministic Metrics} URAF is LLM-judged, but it does not stand alone in the framework. Recompilability is grounded in the compiler (a deterministic oracle: code either compiles and links or does not), and functionality is grounded in the runtime (a deterministic oracle: register/I/O traces either match or diverge). Because the three pipelines are computed independently from different signals, their joint behavior provides built-in cross-validation: if URAF scores were dominated by judge noise rather than real readability differences, the URAF ranking would have to be uncorrelated with the deterministic recompilability and functionality rankings. As we will show in \cref{secfinding}, the three rankings are in fact tightly coupled (IDA $>$ Ghidra $>$ BinaryAI $>$ RetDec/Angr is preserved across all three pipelines and across all three judges), which is the strongest empirical evidence that URAF measures a real signal rather than judge artifact. We therefore use URAF as the diagnostic decomposition (where does the readability gap come from?) and the deterministic metrics as the cross-check on its calibration.

\subsubsection{Recompilability}
Conventional decompiler benchmarks typically treat recompilability as a binary property: the recovered code either compiles or fails. This dichotomy is insufficient for scientific evaluation because it obscures the severity of compilation failures. A function that only misses a single header file is fundamentally different from one containing pervasive structural syntax violations, yet a binary metric treats both cases identically. To obtain a more discriminative measure, we design an LLM-driven Iterative Repair pipeline that estimates the “cost” of transforming raw decompiler output into a valid, linkable binary under a fixed repair budget.

Our approach is implemented as a Python-based automated repair agent (evaluator/syntactic) that forms a closed feedback loop between the compiler and LLM. The agent invokes GCC through a wrapper and, upon failure, parses the compiler’s unstructured diagnostics via an \texttt{ErrorParser} into structured JSON objects that record line numbers, normalized error categories (\eg, Undeclared Identifier, Unknown Type), and raw messages. Using this structured error stack and the problematic code region, the agent constructs a context-rich prompt and delegates remediation to the LLM.
The LLM edits the multi-line structure via \texttt{edit\_code\_block} and uses \texttt{replace\_string} for localized fixes.

Beyond syntax fixes, the agent also handles common decompiler artifacts (e.g., conflicting runtime stubs, unresolved external symbols) by generating minimal stubs that enable linking without changing high-level logic. After each repair, the code is recompiled and the diagnostics are fed back into the loop until a linkable binary is produced or a fixed budget of 50 repair iterations is reached.

\begin{figure}[!th]
  \centering
  \includegraphics[width=0.6\linewidth]{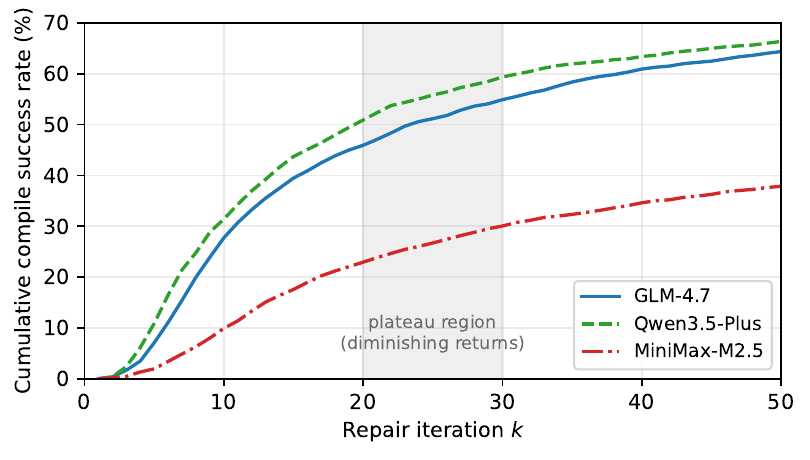}
  \caption{Cumulative compile-success rate vs.\ repair iteration over the full 9,066-trace corpus. The plateau region (20--30 iterations, shaded) bounds where additional budget yields diminishing returns; the 50-iteration cap is in the post-plateau regime for all three LLMs.}
  \label{fig:repair_iter_curve}
\end{figure}

Two design choices warrant explanation. 
First, we use raw iteration count only as a budget cap, 
not as the primary cost metric. The cap of 50 is empirically 
justified by the cumulative compile-success curve in \cref{fig:repair_iter_curve}, 
computed from all 9,066 repair traces: each LLM gains $\sim$20--21 percentage points 
in the first 5--10 iterations, an additional $\sim$11--12 points in iterations 10--15, 
but only $\sim$1.5--1.9 points 
per 5-iteration window beyond iteration 30; 
the curve is essentially flat in the 30--50 region. 
The \(>30\)-iteration tail is therefore informative about repair 
difficulty but not about whether more budget would help---runs 
that exhaust the cap fall into one of two failure modes 
we manually verified on a sample of trace logs, 
namely \emph{oscillating} repair 
(re-introducing an error fixed two iterations earlier) or 
\emph{stuck} states with zero error-count progress; 
both indicate the model has reached its repair ceiling. 
Second, the headline recompilability metric is the three-tier outcome (FS/LF/CF), 
which is invariant to the exact cap once the cap is in the post-plateau regime: 
rerunning the analysis with a 30-iteration cap re-classifies $<2\%$ of tasks. 
To address the concern that iteration count alone is a noisy proxy for repair cost, we additionally report a normalized cost view: for each failed task we compute the fraction of initial compile/linker errors removed before exhaustion (\cref{fig:compilability_failure_effort_ratio}), which captures partial convergence even when the final outcome is CF or LF.

Instead of a simple pass/fail, we categorize outcomes into three tiers based on where the pipeline stalls, capturing the nature and severity of failure:
\circled{1} Full Success (FS):The code compiles and links into a valid binary within 50 iterations, enabling subsequent semantic verification.
\circled{2} Link Fail (LF): Compilation succeeds but linking still fails at budget exhaustion due to unresolved symbols or conflicting definitions, indicating syntactically correct code but an unsatisfied external interface.
\circled{3} Compile Fail (CF): Compilation errors remain when the budget is exhausted, indicating structural or type-level issues in the decompiler output that the LLM cannot repair to a syntactically valid state.

\subsubsection{Functionality}
Our goal in recompilation is semantic fidelity,
meaning that the recompiled (decompiled + rebuilt) program should behave like the original program in execution.
Purely static analysis is not sufficient for this, because it cannot fully resolve pointer aliasing or reconstruct concrete runtime memory state. To address these limitations, we build a Dynamic Semantic Verification pipeline on top of the Frida dynamic instrumentation framework, allowing us to observe and compare runtime behavior at multiple granularities.

\myparagraph{Why Differential Tracing Rather Than Unit Tests}
A natural alternative is to attach an external unit-test suite to the decompiled function and check input--output equivalence. We deliberately use \emph{differential dynamic tracing} for three reasons specific to our setting. (i) Decompilers do not preserve the original calling convention, signature precision, or even parameter count when type recovery fails (e.g., Angr's signed-vs-unsigned drift in \cref{fig:motivation}); a unit-test harness that calls the decompiled function with typed inputs would either silently coerce the inputs (hiding the very type-recovery failures we want to detect) or fail to compile due to the same type drift, making the metric meaningless. Differential tracing observes the original and recompiled binaries on the \emph{same driver} and the \emph{same lowering of inputs}, so type-recovery failures show up as register/I/O divergence rather than as harness errors. (ii) Many decompiler defects manifest only as side effects (a write to stdout, a global state change, a wrong stack adjustment that propagates) rather than as wrong return values; unit tests on returns alone would miss them, while register-superset capture and side-effect attribution to the call frame catch them. (iii) For \texttt{void} functions and functions whose effect is non-local, unit testing has no natural ``expected output''---differential tracing avoids this by comparing the two binaries' traces directly, with the original binary serving as the oracle. We therefore view unit testing as complementary (a strong baseline for individually-correctable functions) rather than as a substitute, and leave a head-to-head comparison---e.g., a mutation-testing experiment that injects controlled semantic faults and measures detection rate of differential tracing vs.\ harness-driven unit tests---as future work.

The core of our infrastructure is a Frida-based hooking agent, injected into the running process as a custom JavaScript module. Unlike emulation-based approaches (e.g., QEMU), this in-process instrumentation is relatively lightweight and transparent. Because real-world binaries are often stripped, we cannot rely on symbol names; instead, we first run \texttt{nm} to extract the relative offsets of all text-segment functions, and at runtime compute absolute entry points as \texttt{BaseAddress + Offset} to attach hooks. This enables robust instrumentation of non-symbolic binaries. To handle recursion and multi-threading, the agent maintains a thread-local call stack: on each function entry it generates a globally unique sequence ID and pushes it; on exit, it pops the corresponding ID. This guarantees that arguments, intermediate state, and return values are consistently attributed to the correct call frame.

The pipeline goes beyond simple return-value tracing to uncover deeper data-flow and side-effect mismatches. BinDeBench uses a deep capture strategy that records an extended set of general-purpose registers (e.g., x0–x7 on ARM64), not just the standard argument registers. This richer snapshot helps detect subtle divergences, such as parameters passed through non-canonical registers or violations of callee-saved conventions. In parallel, we capture side effects by hooking the kernel’s \texttt{write} system call and selectively monitoring \texttt{fd = 1} (\texttt{stdout}). Each captured output string (e.g., Login Failed) is associated with the function currently on the thread-local stack, so I/O behavior can be linked to specific call frames. This allows us to semantically validate even \texttt{void} functions with no explicit return value, based on their observable effects.

We evaluate functionality through multi-level dynamic tracing comparisons, instrumenting the original and recompiled binaries and then comparing their execution traces. \circled{1} Program Level: We parse the standard output of each case into observables and compare the consistency of these values between the two binaries.
\circled{2} Function Level: We perform function-level instrumentation on the binaries to compare the consistency of register values captured at function entry/exit points and return value pairs.
\circled{3} Instruction Level: We trace the execution sequence of instructions, recording the called function names, register signatures, and matching function return value categories to construct normalized labeled sequences, and calculate similarity scores for these sequences.

\myparagraph{Coverage and Scope of Dynamic Verification}
Differential dynamic tracing is, by construction, an \emph{observational} method: its judgement is bounded by the execution paths the case driver actually exercises. To make this property explicit, every test case in \proj is paired with an in-suite driver designed to exercise the function under test directly (rather than relying on incidental paths through a larger program), so the metric reports behavioral agreement on the path the benchmark intends to test, not on the function in full generality. We empirically verified that the in-suite drivers cover the cases they target: parsing the stdout traces of the 567 original binaries (the binaries used as the equivalence oracle), the driver invokes \textbf{95.5\%--100\%} of the expected functions per dimension---Control Flow 97.8\%, Data Types \& Variables 97.9\%, Memory Operations 95.5\%, Function Calls 100\%, Object-Oriented C++ 100\%, Compile-Time Specialization 95.6\%, System Interaction 100\%, Special Challenges 90.8\%. Dimension-level driver coverage is therefore essentially saturated, and the residual is dominated by case-specific environmental dependencies (e.g., one \textit{Special Challenges} case requires a kernel feature not present in the Lima ARM32 image) rather than by structural driver gaps. Three concrete consequences follow. (i) For functions whose behavior is fully captured by their return value or by I/O, our hooks observe the divergence directly. (ii) For \texttt{void} functions and functions with no return-value side effects, we rely on the side-effect channel (hooked \texttt{write} on \texttt{fd=1} associated with the call frame on the thread-local stack), which is why benchmarks are written to print a function-identifying token on the success path; functions for which neither register-level nor side-effect evidence is available are scored as \texttt{Unsupported} rather than as a positive match. (iii) Branches that the driver does not reach are silently \emph{not} disconfirmed by our metric, which is why we report function-level and instruction-level agreement as \emph{conditional} signals on top of the program-level outcome rather than as a substitute for it. The interpretation of high function-level I/O agreement together with low program-level agreement, discussed in \cref{secfinding}, is precisely a manifestation of this conditional structure, not a measurement artifact.

\section{Findings}\label{secfinding}

\subsection{Study Setup}

\myparagraph{Benchmark Construction}
% To construct a comprehensive ground truth dataset, we utilized the 8 source scenarios defined in Section 4.1 (Control Flow, Variable/Type, Memory Ops, Function Call, C++ Features, System Interaction, etc.). We systematically combined these sources with a matrix of compilation options to generate a diverse set of binaries. The configuration space includes:
% \begin{itemize}
%     \item \textbf{Compilers (2)}: GCC and Clang.
%     \item \textbf{Optimization Levels (5)}: O0, O1, O2, O3, and Os.
%     \item \textbf{Debug Information (2)}: With symbols (\texttt{-g}) and stripped (\texttt{no\_g}).
%     \item \textbf{Architectures (6)}: x86, x64, ARM32, ARM64, MIPS32, and MIPS64.
% \end{itemize}
% This combinatorial expansion results in a total of $8 \times 2 \times 5 \times 2 \times 6 = 960$ distinct binary executables. We utilized \texttt{Podman} to containerize the build environment, creating a dedicated Docker image pre-installed with the full suite of cross-compilation toolchains (e.g., \texttt{gcc-multilib}, \texttt{binutils-multiarch}). A batch automation script was deployed to manage the compilation of all 960 targets, ensuring reproducibility and consistency.
% To build a comprehensive dataset, we combined the design and build settings from~\cref{subsecbench} to maximize binary diversity. Specifically, this included the GCC and Clang compilers; optimization levels (-O0, -O1, -O2, -O3, and -Os); dependency on debug information; and x86, x64, ARM32, ARM64, MIPS32, and MIPS64 architectures. Ultimately, the dataset contained 960 different executable binaries.
To construct a comprehensive benchmark, we combined the design and build settings described in \cref{subsecbench} so as to maximize binary diversity. The 240 atomic test functions are grouped into \textbf{8 dimensional source files} (one per dimension in \cref{tab:bench}); each source file is then compiled along five orthogonal axes:
\begin{itemize}[leftmargin=10pt,topsep=0pt,itemsep=0pt,parsep=0pt]
    \item compiler $\in$ \{GCC, Clang\} (2 choices),
    \item optimization $\in$ \{\texttt{-O0}, \texttt{-O1}, \texttt{-O2}, \texttt{-O3}, \texttt{-Os}\} (5 choices),
    \item debug symbols $\in$ \{with \texttt{-g}, stripped\} (2 choices),
    \item architecture $\in$ \{x86, x64, ARM32, ARM64\} (4 choices).
\end{itemize}
This yields exactly $8 \times 2 \times 5 \times 2 \times 4 = 640$ distinct executable binaries. We deliberately compile at the file-per-dimension granularity rather than the function-per-binary granularity to keep the configuration matrix tractable while still spanning 240 atomic functions inside those 640 binaries; per-dimension grouping also gives us linkable, self-contained executables for end-to-end recompilability and dynamic verification (see \cref{subsecbench}).

\myparagraph{Decompiler Selection}
 We evaluated five mainstream whole-program decompilers: IDA Pro~\cite{ida}, 
 Ghidra~\cite{ghidra}, RetDec~\cite{retdec}, Angr~\cite{shoshitaishvili2016state}, 
 and BinaryAI~\cite{binaryai}. 
 These tools share a common interface---they accept an executable and emit recompilable, 
 function-organized C---which is the precondition 
 for our recompilability and program-level functionality pipelines. 
 LLM-based decompilers (LLM4Decompile~\cite{tan2024llm4decompile} and 
 DecLLM~\cite{wong2025decllm}) operate at function granularity and do not produce a linkable program; they are therefore not directly comparable on the program-level recompilability metric. URAF and the function-level functionality metric, however, can in principle be applied at function granularity, and we include LLM-based decompilers on those two dimensions as a separate function-level study (see \cref{secfinding}); the program-level numbers in \cref{tab:global_compilability} and \cref{tab:semantic_overview_glm} continue to reflect only whole-program decompilers, to avoid mixing comparable and non-comparable units.

\myparagraph{LLM Setup} The pipeline uses LLMs in two roles: (i) as a \emph{judge} for URAF readability scoring, and (ii) as a \emph{repair agent} for the recompilability pipeline. We use three models in both roles to control for judge/repair-model bias: GLM-4.7, Qwen3.5-Plus, and MiniMax-M2.5. All three are accessed via their respective hosted chat-completion APIs; none of the experiments rely on local fine-tuning or model-specific finetunes. Decoding parameters are fixed across runs: temperature 0, top-$p$ 1.0, no repetition penalty, no system-side tool calling other than our own \texttt{edit\_code\_block} and \texttt{replace\_string} primitives in the repair agent. Each readability evaluation issues one prompt of the form documented in \cref{tab:readiablity}; each recompilability run is capped at 50 repair iterations as discussed in \cref{subsecbench}; functionality runs use no LLM. \cref{tab:efficiency_summary} reports the resulting end-to-end token consumption and wall-clock cost across the full 640-binary $\times$ 5-decompiler $\times$ 3-LLM matrix.

% M2M, and LLM4decompile [35]. 

% \circled{6} LLM4decompile~\cite{tan2024llm4decompile}, 

% We evaluated 7 decompilers against this dataset: IDA Pro, Ghidra, RetDec, Angr, LLM4decompile, BinaryAI, and M2M. The decompilation results are as follows:
% \begin{itemize}
%     \item \textbf{Full Success (960/960)}: IDA Pro, Ghidra, BinaryAI, and M2M successfully generated pseudo-C output for all 960 binaries.
%     \item \textbf{Partial Success}:
%     \begin{itemize}
%         \item \textbf{RetDec}: Failed to decompile all MIPS architecture binaries due to lack of support.
%         \item \textbf{Angr}: Failed to process the C++ test cases.
%         \item \textbf{LLM4decompile}: Operated in both 'End-to-End' and 'Refine' modes. Since this model functions at the sub-routine level, we implemented a custom stitching script to assemble the full decompiled file from individual function outputs, focusing specifically on the \texttt{main} function and our custom test functions.
%     \end{itemize}
% \end{itemize}

\begin{table*}[!t]
\centering
\caption{Readability Overview.
\colorbox{gray!35}{Dark} and \colorbox{gray!10}{light} cells represent the highest and lowest average scores, respectively.}
% \colorbox{gray!35}{Dark shaded cells} indicate the highest average score, while \colorbox{gray!10}{light shaded cells} indicates the lowest.}
\label{tab:global_readability}
\resizebox{\linewidth}{!}{%
\begin{tabular}{l c c c c c}
\toprule
\textbf{Decompiler} & \textbf{Lexical} & \textbf{Structural} & \textbf{Type} & \textbf{Semantic} & \textbf{Contextual} \\
                    & Avg. (GLM/Qwen/MiniMax) & Avg. (GLM/Qwen/MiniMax) & Avg. (GLM/Qwen/MiniMax) & Avg. (GLM/Qwen/MiniMax) & Avg. (GLM/Qwen/MiniMax) \\
\midrule
\textbf{IDA} & \cellcolor{gray!35} 5.89 (6.58/5.31/5.78) & \cellcolor{gray!35} 7.26 (7.93/6.37/7.48) & \cellcolor{gray!35} 6.13 (7.06/5.10/6.22) & 3.46 (3.57/3.20/3.59) & \cellcolor{gray!35} 5.92 (7.00/4.95/5.82) \\
\textbf{Ghidra} & 5.44 (5.96/5.18/5.18) & 6.88 (7.49/6.08/7.07) & 5.50 (6.04/4.94/5.53) & \cellcolor{gray!35} 3.99 (4.10/3.76/4.10) & 5.70 (6.46/5.08/5.56) \\
\textbf{BinaryAI} & 5.38 (6.02/4.92/5.20) & 6.90 (7.71/5.90/7.10) & 4.67 (5.17/4.06/4.78) & 2.49 (2.61/2.39/2.47) & 5.52 (6.35/4.84/5.37) \\
\textbf{RetDec} & 4.59 (4.97/4.38/4.42) & 5.84 (6.24/5.06/6.23) & 4.35 (4.53/3.74/4.78) & 2.47 (2.42/2.29/2.71) & 5.32 (5.66/4.87/5.44) \\
\textbf{Angr} & \cellcolor{gray!10} 4.45 (4.90/4.13/4.30) & \cellcolor{gray!10} 5.76 (6.23/4.88/6.18) & \cellcolor{gray!10} 4.35 (4.66/3.77/4.61) & \cellcolor{gray!10} 2.47 (2.45/2.41/2.54) & \cellcolor{gray!10} 4.77 (5.25/4.26/4.79) \\ 
\bottomrule
\end{tabular}
}
\end{table*}
\myparagraph{Pipeline Implementation}
Our evaluation framework comprises three sequential phases, each addressing a distinct decompilation objective:
\textit{\circled{1} Readability}.
We interact with the LLM API through the OpenAI Python library to enforce text output with rule constraints and strict JSON pattern output extracted by deterministic regular expressions, ensuring deterministic parsing and reproducible results.
\textit{\circled{2} Recompilability}. Our auto-repair agent consumes machine-readable diagnostic messages from GCC/Clang to identify compilation errors. To ensure experimental fairness, we mandate that all repair attempts use identical compilation commands to those employed during original binary construction.
\textit{\circled{3} Functionality}. We employ Frida-based~\cite{frida} dynamic instrumentation for fine-grained semantic verification. The system automatically dispatches verification tasks to Lima VM~\cite{lima} instances provisioned with matching target architectures. During execution, Frida injects our JavaScript agent to intercept function invocations and write system calls, computing fidelity metrics in real time.

% \textbf{Readability}: We utilize the \texttt{openai} Python library to interface with the LLM API, enforcing strict JSON schema output for parsing stability.

%  \textbf{Recompilability}: The automated repair agent detects errors using \texttt{gcc/clang}'s machine-readable diagnostic output. Crucially, during the repair process, we enforce the exact same compilation command used to build the original binary, ensuring a fair baseline.
 
% \textbf{Functionality}: Dynamic instrumentation is powered by \texttt{Frida}. The pipeline automatically dispatches the verification task to the appropriate Lima VM instance matching the binary's architecture. \texttt{Frida} injects our JavaScript agent to intercept function execution and `write` syscalls, calculating fidelity metrics in real-time.

\myparagraph{Evaluation Environment} All experiments ran on an Apple M4 Pro (ARM64) workstation with 64\,GB RAM. To support heterogeneous binary execution, we deployed a Lima-managed virtualization cluster hosting architecture-specific Linux environments. LLM decoding parameters are listed in the \textit{LLM Setup} paragraph above and held fixed throughout.

% To support the execution of heterogeneous binaries, we established a virtualized cluster using Lima to manage distinct Linux environments. 

% ARM64 Native with Ubuntu 24.04,
% x64 Emulated via Rosetta 2, 
% ARM32 Full Emulation with Ubuntu 18.04, and 86 Full Emulation, Ubuntu 18.04. 

% \textit{Note}: Due to the technical limitation of lacking a MIPS-compatible Lima instance on the host, the \textbf{Semantic Evaluation (Step 3)} was skipped for MIPS binaries, although Readability and Recompilability were still assessed.

\subsection{Readability}

\cref{tab:global_readability} summarizes the overall readability results and reveals a clear and consistent ranking among decompilers. IDA delivers the best overall performance, achieving the highest scores in \textit{Lexical Clarity}, \textit{Structural Intelligibility}, \textit{Type-System Fidelity}, and \textit{Contextual Coherence}. Ghidra leads on \textit{Semantic Transparency}, indicating a relative advantage in preserving high-level semantic cues such as idioms, annotations, and exceptional control flow. Angr consistently ranks last on all dimensions, suggesting that, despite its strengths in binary analysis and symbolic execution, it is significantly weaker at reconstructing high-level readability and semantic structure. The interesting finding is that \textit{Semantic Transparency} emerged as a universal shortcoming for all tools: even though top-tier tools generate code that is sufficiently clean in syntax and structure (with higher \textit{Structural Intelligibility} scores), a significant gap remains in \emph{decompiler-emitted} annotations (comments produced by the decompiler from idiom recognition, signature databases, or debug info), recognized idioms, and indications of deep intent. This is distinct from "preserving the original comments," which is impossible because compilation strips them; the \textit{Semantic Transparency} measurement targets the decompiler's ability to \emph{reconstruct} such annotations from binary evidence. The result indicates that making code look clean is not equivalent to making it easy to understand.

The three LLM-based evaluators exhibit highly consistent trends: GLM-4.7 tends to assign slightly higher absolute scores, while Qwen3.5-Plus is stricter, but these scale differences do not affect their relative judgements. To quantify cross-judge agreement we compute two complementary statistics on the 25 (decompiler, level) cells of \cref{tab:global_readability}. (i) \emph{Absolute spread:} the standard deviation of the three judges' mean scores within a cell has median 0.56 and maximum 1.03 across the 25 cells; 24 of 25 cells stay within $\pm$1 point on the 10-point scale, the single exception being IDA's \textit{Contextual Coherence}. (ii) \emph{Rank agreement:} for each of the five readability levels we rank the five decompilers under each judge and compute pairwise Spearman $\rho$. The mean pairwise $\rho$ is 0.90 (GLM--Qwen), 0.88 (GLM--MiniMax), and 0.86 (Qwen--MiniMax), and the per-level $\rho$ is $\geq 0.8$ on four of the five levels; the lone outlier is \textit{Semantic Transparency} ($\rho$ as low as 0.6 between GLM and MiniMax) which sits at the bottom of the absolute scale for every decompiler and is therefore the most sensitive to absolute-scale drift. The exact top-5 ordering (IDA $>$ Ghidra or BinaryAI $>$ \ldots $>$ Angr) shifts in adjacent neighbor swaps (BinaryAI vs.\ Ghidra at \textit{Lexical Clarity} and \textit{Structural Intelligibility}; Angr vs.\ RetDec at \textit{Type-System Fidelity}); the head-to-tail ordering (IDA at the top, Angr/RetDec at the bottom) is unanimous on all five levels. We therefore report cross-judge means in figures throughout, treating the cross-judge spread as inter-rater noise; per-judge values remain in the tables for transparency.

\finding{Even the best decompiler scores only 6.43/10 on average URAF readability, and the universal worst dimension across \emph{every} decompiler and \emph{every} judge is \textit{Semantic Transparency} (2.5--4.0/10)---decompilers can produce structurally clean code, but they cannot recover \emph{intent}. This bottleneck is fundamental, not tool-specific.}

% Overview of table 1.
% (1) IDA shows the best overall performance, achieving the highest scores in lexical clarity (L1), structural interpretability (L2), type system recovery (L3), and contextual coherence (L5).
% (2) Ghidra ranks first in semantic transparency (L4), which are higher than other decompilers. This indirectly confirms its distinctive advantage in handling constructs such as idioms, annotations, and exceptional logic.
% (3) angr ranks last across all evaluated dimensions. This suggests that although it is powerful in binary analysis and symbolic execution, it performs poorly in recovering high-level readability and semantic structure.
% (4) The evaluations produced by the three LLMs exhibit consistent trends, although their absolute mean scores differ slightly. Among them, GLM-4.7 appears to assign relatively higher scores, while Qwen3.5-Plus is comparatively the most stringent. This discrepancy is likely due to minor absolute-value biases in how different models respond to the prompt. Nevertheless, their relative assessments—particularly the ability to distinguish the quality of different decompilers—remain highly consistent.

\subsubsection{Single-Factor Effects}  
We analyze the impact of individual variables on readability. To save space, the scores are expressed as the average of the three LLMs.
If there is no significant indication of per-level scores, then it is the average score across the five levels.

\begin{table}[!ht]
\centering
\caption{Readability comparison under different compilers.}
\label{tab:compiler_sensitivity}
\resizebox{0.7\columnwidth}{!}{%
\begin{tabular}{l|ccccc|ccccc}
\toprule
& \multicolumn{5}{c|}{\textbf{GCC}} & \multicolumn{5}{c}{\textbf{Clang}} \\
\textbf{Decompiler} & \textbf{Lex} & \textbf{Str} & \textbf{Typ} & \textbf{Sem} & \textbf{Ctx} & \textbf{Lex} & \textbf{Str} & \textbf{Typ} & \textbf{Sem} & \textbf{Ctx} \\
\midrule
IDA      & \cellcolor{gray!35}5.93 & \cellcolor{gray!35}7.32 & \cellcolor{gray!35}6.17 & 3.53 & \cellcolor{gray!35}5.91 & \cellcolor{gray!35}5.85 & \cellcolor{gray!35}7.20 & \cellcolor{gray!35}6.08 & 3.38 & \cellcolor{gray!35}5.93 \\
Ghidra   & 5.48 & 6.97 & 5.60 & \cellcolor{gray!35}4.10 & 5.73 & 5.40 & 6.79 & 5.41 & \cellcolor{gray!35}3.87 & 5.68 \\
BinaryAI & 5.38 & 6.96 & 4.65 & 2.53 & 5.44 & 5.38 & 6.84 & 4.68 & 2.44 & 5.61 \\
RetDec   & 4.59 & 5.77 & \cellcolor{gray!10}4.38 & \cellcolor{gray!10}2.52 & 5.32 & 4.60 & 5.97 & \cellcolor{gray!10}4.29 & \cellcolor{gray!10}2.38 & 5.32 \\
Angr     & \cellcolor{gray!10}4.42 & \cellcolor{gray!10}5.77 & 4.33 & 2.47 & \cellcolor{gray!10}4.71 & \cellcolor{gray!10}4.47 & \cellcolor{gray!10}5.76 & 4.36 & 2.46 & \cellcolor{gray!10}4.82 \\
\bottomrule
\end{tabular}%
}
\end{table}

\myparagraph{Compiler} \cref{tab:compiler_sensitivity} compares the readability scores of decompilers under GCC and Clang. For most decompilers, the differences between compilers are narrow. In contrast, the differences between different decompilers are much larger. This suggests that the compiler has a relatively small impact on decompilation.

\begin{figure}[!ht]
  \centering
  \includegraphics[width=0.6\linewidth]{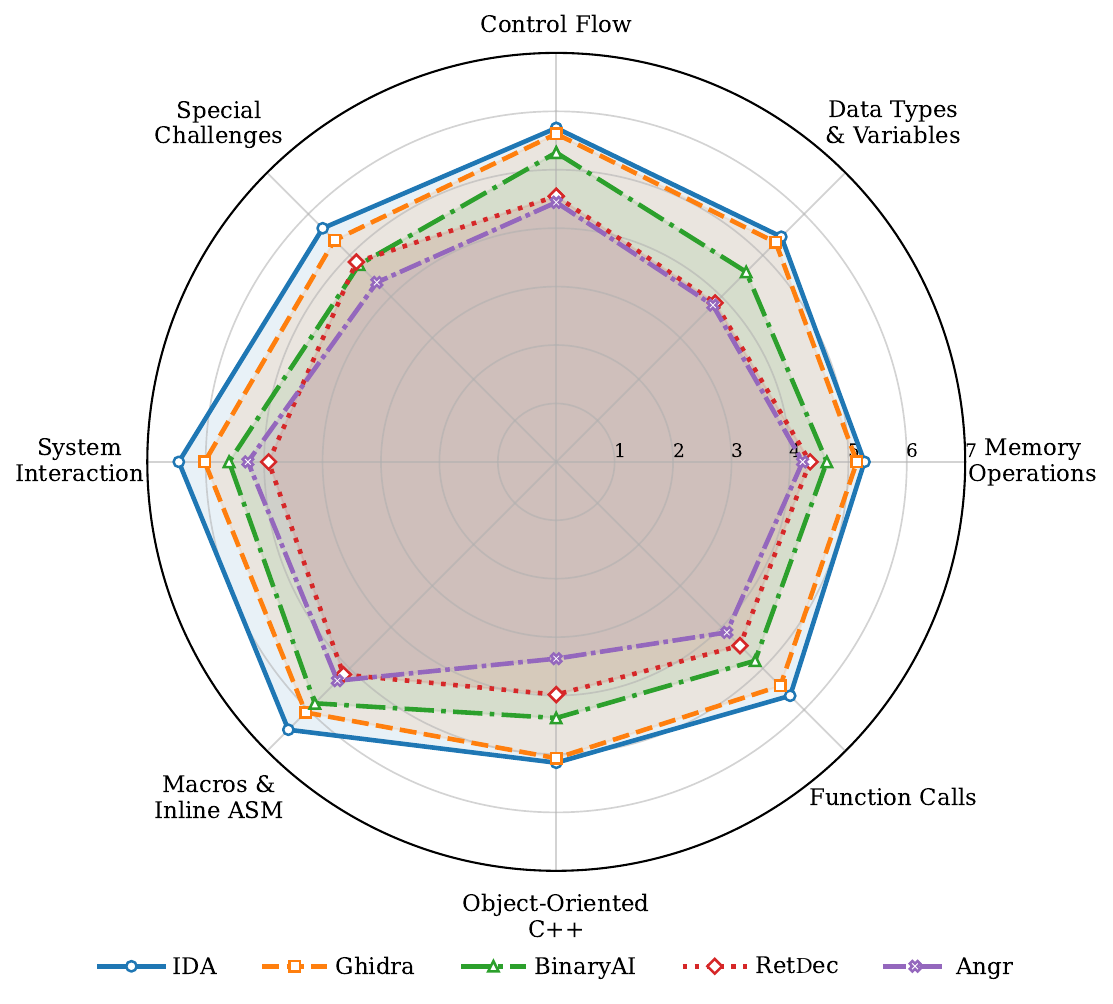}
  \caption{Readability across eight dimensions.}
  \label{fig:decompiler_readability_radar}
\end{figure}

\myparagraph{Dimension} \cref{fig:decompiler_readability_radar} illustrates the scores across different dimensions of the test cases.
IDA achieved the best overall performance, with a particularly notable advantage in challenging dimensions such as Macros \& Inline Assembly and System Interaction. Furthermore, Object-Oriented C++ was identified as the generally most difficult dimension, posing a major challenge especially for Angr and RetDec, while the rankings among the tools remained highly stable across dimensions.

% \begin{table}[t]
% \centering
% \caption{Readability across architectures.}
% \label{tab:readability_by_arch}
% \resizebox{0.7\columnwidth}{!}{%
% \begin{tabular}{lccccc}
% \toprule
% \textbf{Decompiler} & \textbf{arm32} & \textbf{arm64} & \textbf{x86} & \textbf{x64} & \textbf{Avg.} \\
% \midrule
% IDA      & \cellcolor{green!15}5.68 & \cellcolor{green!15}5.79 & \cellcolor{green!15}5.77 & \cellcolor{green!15}5.69 & \cellcolor{green!15}5.73 \\
% Ghidra   & 5.49 & 5.58 & 5.44 & 5.49 & 5.50 \\
% BinaryAI & 4.97 & 4.96 & 4.87 & 5.16 & 4.99 \\
% Retdec   & 4.50 & \cellcolor{red!15}4.29 & 4.58 & 4.61 & 4.50 \\
% Angr     & \cellcolor{red!15}4.18 & 4.43 & \cellcolor{red!15}4.25 & \cellcolor{red!15}4.57 & \cellcolor{red!15}4.36 \\
% \bottomrule
% \end{tabular}%
% }
% \end{table}

\myparagraph{Architecture}  Architecture introduces only mild fluctuations and does not change the overall ranking. Weaker tools are actually more sensitive to architecture. IDA and Ghidra perform robustly across architectures, while Angr, RetDec, and even BinaryAI are more volatile, the easier it is to be stumped by specific architectures.

\begin{figure}[!ht]
  \centering
  \includegraphics[width=0.7\linewidth]{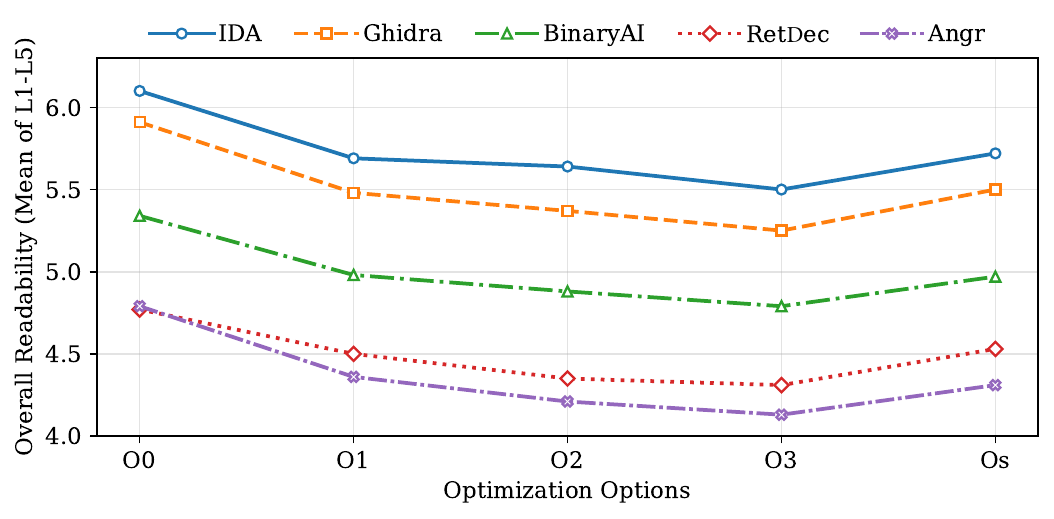}
  \caption{Overall readability across optimization options.}
  \label{fig:readability_by_opt}
\end{figure}

\myparagraph{Optimization} 
\cref{fig:readability_by_opt} demonstrates that aggressive compiler optimization significantly degrades decompiled readability, with \texttt{O0} yielding the best results and \texttt{O3} the worst. However, an interesting finding is that \texttt{Os} (size optimization) outperforms \texttt{O2} and \texttt{O3}, often rivaling \texttt{O1}. This suggests that optimizing for code size is less detrimental to readability than aggressive performance tuning, likely because \texttt{Os} avoids the severe control-flow restructuring and complex transformations typical of high-level optimizations.

\begin{table}[!ht]
\centering
\caption{Impact of debug symbol on readability. Each value reports the improvement: $\Delta=\texttt{with-debug}-\texttt{without-debug}$. }
\label{tab:debug_info_delta}
\resizebox{0.6\columnwidth}{!}{%
\begin{tabular}{lcccccc}
\toprule
\textbf{Decompiler} & \textbf{$\Delta$Lex} & \textbf{$\Delta$Str} & \textbf{$\Delta$Typ} & \textbf{$\Delta$Sem} & \textbf{$\Delta$Ctx} & \textbf{$\Delta$Avg.} \\
\midrule
IDA      & \cellcolor{gray!35}0.52 & \cellcolor{gray!35}0.17 & \cellcolor{gray!35}1.69 & 0.45 & \cellcolor{gray!35}0.39 & \cellcolor{gray!35}0.64 \\
Ghidra   & 0.18 & \cellcolor{gray!10}0.00 & 1.58 & 0.49 & \cellcolor{gray!35}0.39 & 0.53 \\
BinaryAI & 0.01 & 0.06 & 0.06 & \cellcolor{gray!10}0.02 & 0.02 & 0.03 \\
RetDec   & 0.21 & 0.11 & 0.47 & \cellcolor{gray!35}0.56 & 0.27 & 0.32 \\
Angr     & \cellcolor{gray!10}-0.03 & \cellcolor{gray!10}0.00 & \cellcolor{gray!10}0.02 & \cellcolor{gray!10}0.02 & \cellcolor{gray!10}-0.01 & \cellcolor{gray!10}0.00 \\

\bottomrule
\end{tabular}%
}
\end{table}

\myparagraph{Debug}  
Table \ref{tab:debug_info_delta} quantifies the effect of debug symbols by comparing readability scores with and without \texttt{-g}. IDA benefits the most, with an average gain of +0.64, and these improvements are concentrated mainly in \textit{Type-System Fidelity} and, to a lesser extent, \textit{Semantic Transparency}. This indicates that debug symbols primarily enhance type- and semantics-related aspects by providing explicit type definitions, while having little impact on \textit{Structural Intelligibility}. Since code structure
 is largely determined by the compiled control flow and optimizations rather than by debug metadata, \textit{Structural Intelligibility} remains almost unchanged across all decompilers.

\subsubsection{Multi-Factor Interactions} Building upon the single-factor analysis, we further investigate how the interactions among multiple factors influence the readability of decompiled code, aiming to uncover more implicit, underlying patterns.

\begin{figure}[!ht]
  \centering
  \includegraphics[width=0.8\linewidth]{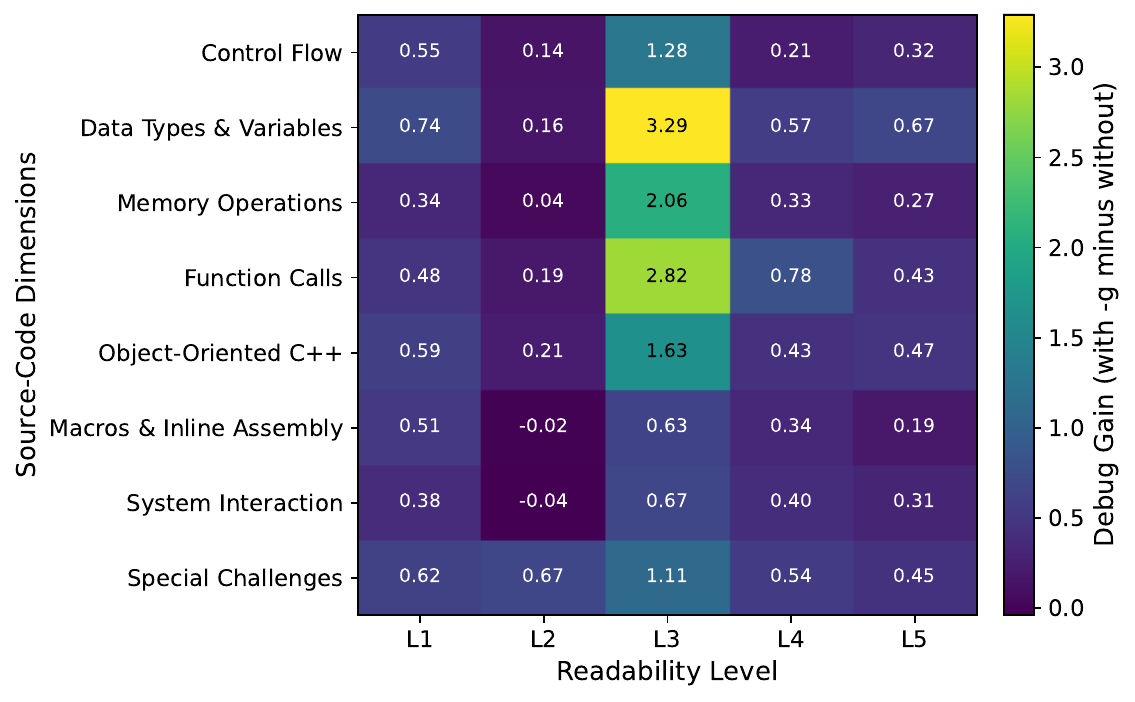}
  \caption{Dimension-specific debug gain of IDA. Each cell reports the improvement brought by compiling with debug.}
  \label{fig:ida_src_debug_heatmap}
\end{figure}

\myparagraphnew{Which dimension with test cases is most significantly affected by debugging symbol} 
\cref{fig:ida_src_debug_heatmap} reveals the differences in the impact of debugging symbol on different code dimensions.
We found the most significant improvement 
in \textit{Type-System Fidelity},
while \textit{Structural Intelligibility} 
barely budges—suggesting debug symbols act like a "type system transplant" rather than a structural makeover. This pattern peaks dramatically in Data Types \& Variables, where \textit{Type-System Fidelity} improvements dwarf all other levels combined.
Moreover, debug symbols prove transformative when decompilers must reconstruct source-level abstractions tethered to metadata (types, function signatures), yet fizzles in dimensions hinging on environmental logic or higher-level semantics.

\begin{figure}[!ht]
  \centering
  \includegraphics[width=0.8\linewidth]{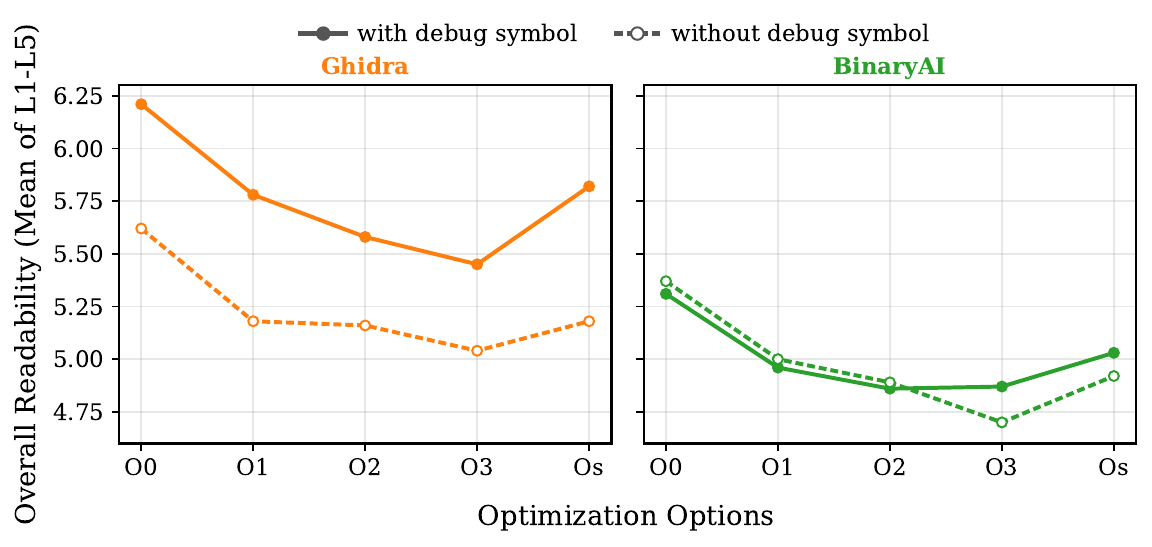}
  \caption{Readability across optimizations \textit{w/wo-debug}.}
  \label{fig:ghidra_binaryai_opt_debug_1x2}
\end{figure}

\myparagraphnew{Differences in the information provided to the decompiler by debug symbols and optimization options} 
\cref{fig:ghidra_binaryai_opt_debug_1x2} reveals an interesting split in how decompilers benefit from debugs under different optimization levels. Tools like Ghidra can meaningfully leverage debug symbols to counteract the loss of readability caused by optimization. For example, Ghidra’s readability at \texttt{O1} with debug symbols (5.78) even surpasses \texttt{O0} without debug symbols (5.62). 
In contrast, BinaryAI fails to use debug symbols in a consistently beneficial way: BinaryAI’s \texttt{O1} + debug score (4.96) remains below \texttt{O0} without debug (5.37), and its small gains at higher optimization levels only barely offset losses at lower ones, yielding a negligible average improvement. 
Collectively, these results highlight a striking difference: some decompilers can turn debug symbols into real readability gains, while others cannot reliably do so.

% \finding{%

% }

\subsection{Recompilability}

\begin{figure}[!t]
  \centering
  \includegraphics[width=\linewidth]{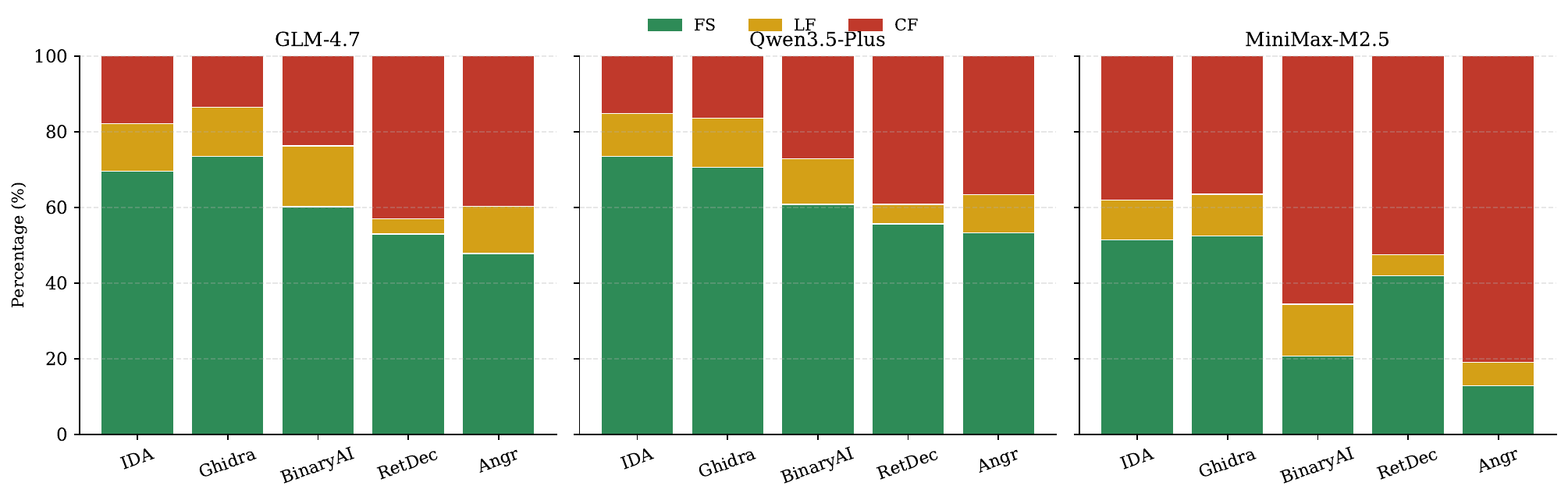}
  \caption{Outcome distribution by decompilers.}
  \label{fig:compilability_by_decompiler}
\end{figure}

\cref{tab:global_compilability} and \cref{fig:compilability_by_decompiler} illustrate the ability to recompile decompiled results for repair at both the LLM and decompiler levels. First, IDA and Ghidra maintain a predominantly successful distribution across all three LLMs. Angr performs the weakest overall, almost degenerating into a state of compilation failure under MiniMax.
Second, compilation failures are the primary cause; link repair failures after successful compilation are relatively reduced, which is related to our test cases maintaining independent external linking.
Finally, Qwen3.5-Plus remains the strongest repair model overall with a success rate of 62.8\%, while MiniMax-M2.5 only achieves 35.8\%. Combined with \cref{tab:global_readability}, it can be seen that decompilers with better readability are often more easily recompiled by LLMs through syntax correction.

\myparagraph{What does the LLM actually fix} To make the repair workload concrete, we mined the per-iteration error categories produced by the \texttt{ErrorParser} across all repair traces. Compile-phase diagnostics dominate (about 83\% of all error instances vs.\ 17\% for the linker phase), and the top categories are: \textit{Syntax Error} (31.0\%), \textit{Undeclared Identifier} (7.6\%), \textit{Conflicting Types} (6.1\%), \textit{Incompatible Pointer Type} (5.6\%), \textit{Type Conversion Warning} (3.3\%), \textit{Implicit Function Declaration} (2.8\%), \textit{Member Access Error} (2.2\%), \textit{Multiple Definition} (2.2\%), \textit{Argument Count Mismatch} (2.2\%), \textit{Void Value Error} (2.1\%), \textit{Unknown Type} (2.1\%), \textit{Undefined Reference} (1.8\%), \textit{Redefinition} (1.8\%), and \textit{Incomplete Type} (1.3\%); the remaining $\sim$26\% are uncategorized residuals. Two patterns matter for decompiler designers. (i) Type-related categories (\textit{Conflicting Types}, \textit{Incompatible Pointer Type}, \textit{Unknown Type}, \textit{Incomplete Type}, \textit{Void Value Error}, \textit{Member Access Error}) together account for $\sim$19\% of all error instances---\emph{type-system collapse is the largest non-syntactic class of repair work}, consistent with \cref{fig:ida_src_debug_heatmap}. (ii) Linker-phase entries (\textit{Multiple Definition}, \textit{Undefined Reference}) are far less frequent in the aggregate but are concentrated in the ARM64 and \textit{Memory Operations} regimes, which is why \cref{fig:compilability_by_arch} shows ARM64 specifically shifting toward link-stage failure. Surface syntax fixes are common but not what determines whether a binary eventually links; type and symbol-resolution errors are what current LLM repair agents struggle with.

\finding{Among 428k repair-error instances mined from the trace logs, type-related categories (\textit{Conflicting Types}, \textit{Incompatible Pointer Type}, \textit{Unknown Type}, \textit{Incomplete Type}, \textit{Member Access Error}, \textit{Void Value Error}) collectively account for $\sim$19\%---comparable to all non-syntax categories combined. \emph{Type-system reconstruction, not surface syntax, is the dominant cost of recompilability repair.}}

\begin{table}[!ht]
\centering
\caption{Recompilability overview. 
Each cell represents the Avg. of LLMs and its percentage, followed by the number of GLM/Qwen/MiniMax respectively.}
\label{tab:global_compilability}
\resizebox{0.9\linewidth}{!}{%
\begin{tabular}{lccc}
\toprule
\textbf{Decompiler} & \textbf{Full Success (FS) } & \textbf{Link Fail (LF)} & \textbf{Compile Fail (CF)}  \\
\midrule
\textbf{IDA} & 414.7 64.8\% (445/470/329)  & 73.3 11.5\% (80/73/67)  & 152 23.8\% (115/97/244)   \\
\textbf{Ghidra} & \cellcolor{gray!35} 419.3 65.5\% (470/452/336)  & 78.7 12.3\% (83/83/70)  &  142 22.2\% (87/105/234)  \\
\textbf{BinaryAI} & 302 47.2\% (385/389/132)  & 89.3 14\% (103/77/88)  & 248.7 38.9\% (152/174/420)  \\
\textbf{RetDec} & 321 50.2\% (339/356/268)  & 31.7 4.9\% (26/33/36)  & 287.3 44.9\% (275/251/336)  \\
\textbf{Angr} & \cellcolor{gray!10} 243 38\% (306/341/82) & 61.3 9.6\% (80/64/40) &  335.7 52.4\% (254/235/518)  \\
\midrule
\textbf{Total} & \textbf{1700 53.1\% (1945/2008/1147)} & \textbf{334.3 10.4\% (372/330/301)} & \textbf{1165.7 36.4\% (883/862/1752)}\\
\bottomrule
\end{tabular}%
}
\end{table}

% \pz{Intersection graph of the three LLMs}

\subsubsection{Single-Factor Effects} Although the performance of different LLMs and decompilers varies, their distribution trends are generally similar. To save space, we select representative ones for analysis.

\begin{figure}[!ht]
\centering
\includegraphics[width=\linewidth]{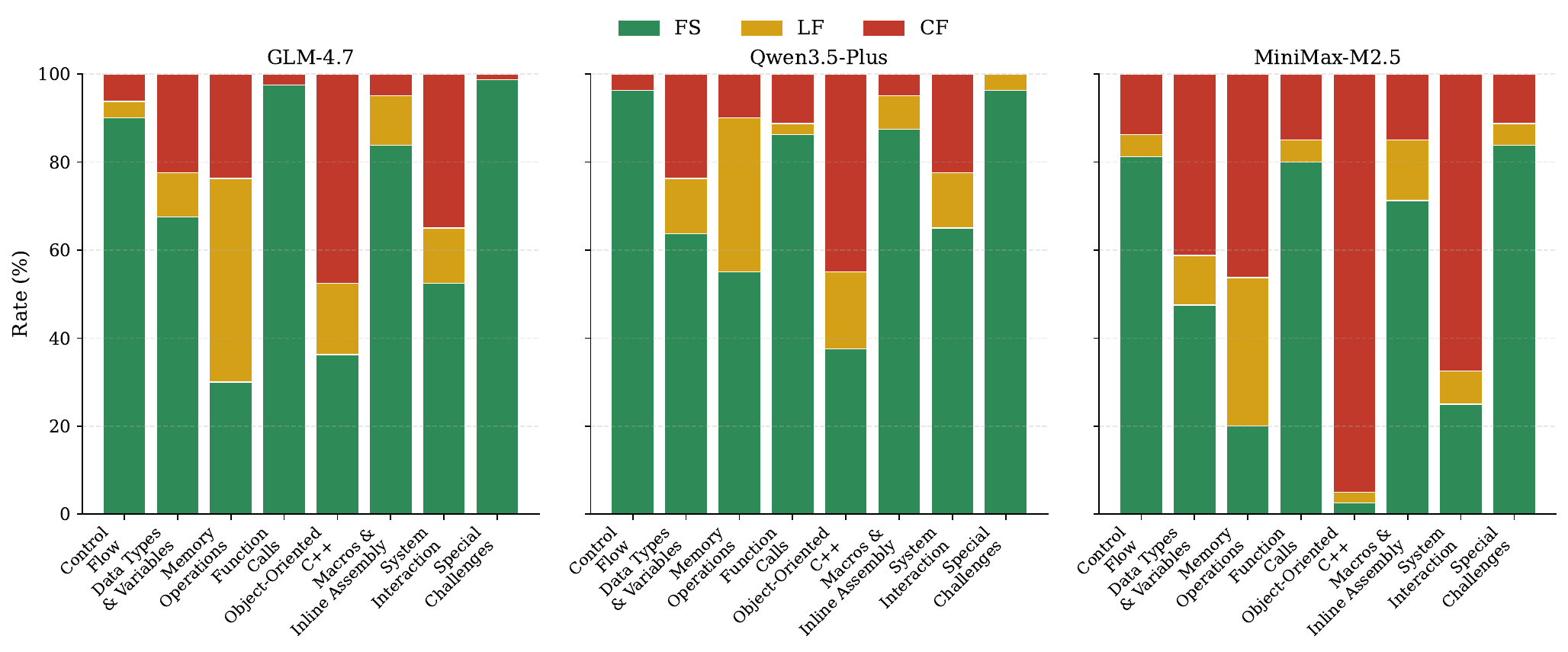}
\caption{IDA-only FS/LF/CF distribution across dimensions. }
\label{fig:compilability_by_src_ida}
\end{figure}

\myparagraph{Dimension} 
% Dimension-level difficulty is even more discriminative than architecture. \cref{tab:compilability_by_src} shows that \textit{Object-Oriented C++} is the hardest dimension for all three LLMs: GLM reaches only 20.3\% compile success, Qwen 28.1\%, and MiniMax 3.1\%, while MiniMax's compile-failed rate rises to 95.8\%. \textit{Memory Operations} forms a second distinctive regime: instead of collapsing directly into compile failure, many tasks stall in linker failure, especially for GLM (43.6\%) and Qwen (35.8\%). This is a qualitatively different difficulty pattern from \textit{Object-Oriented C++}, where compile failure dominates. At the other end of the spectrum, \textit{Function Calls}, \textit{Macros \& Inline Assembly}, and \textit{Special Challenges} are consistently easier. For GLM and Qwen, these dimensions frequently exceed 85\% compile success.
\cref{fig:compilability_by_src_ida} shows the relationship between dimensions and recompilation in the IDA results.
For the three LLMs, memory operation and object-oriented C++ remain the most challenging dimensions. Linking failures caused by memory operation are particularly prominent. This is due to complex reasons such as pointer aliases and object access. This indicates that even after controlling the quality of the decompiler, the dimensionality of the source code still largely determines the difficulty of fixing the problem.

 % \textit{Object-Oriented C++} remains the hardest dimension for all three LLMs: FS is only 36.2\% for GLM, 37.5\% for Qwen, and 2.5\% for MiniMax, while MiniMax's CF rises to 95.0\%. \textit{Memory Operations} forms a distinct LF-heavy regime under IDA, with LF reaching 46.2\% for GLM and 35.0\% for Qwen instead of collapsing primarily into CF. At the other end of the spectrum, \textit{Function Calls} and \textit{Special Challenges} are highly recoverable under IDA: GLM exceeds 97\% FS on both, Qwen exceeds 86\% and 96\%, and even MiniMax stays around 80\% or above. This shows that even after controlling for decompiler quality, source-code dimension still strongly determines whether repair converges to FS, stalls in LF, or falls into CF.

\begin{figure}[!ht]
  \centering
  \includegraphics[width=\linewidth]{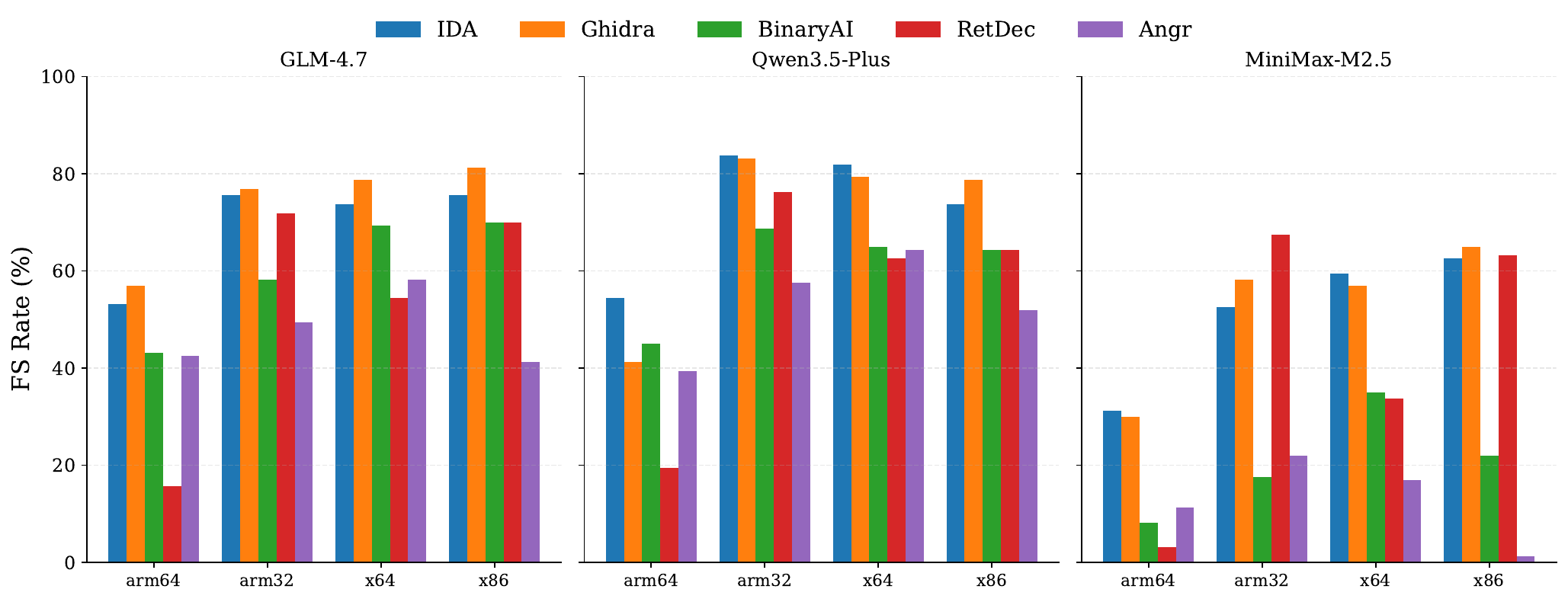}
  \caption{Compile-success rate across target architectures.}
  \label{fig:compilability_by_arch}
\end{figure}

\myparagraph{Architecture}
\cref{fig:compilability_by_arch} shows that recompilation difficulty varies across architectures, with arm64 performing significantly worse than the other three. An analysis of failure causes indicates that arm64 shifts a large proportion of tasks into the linker-failure stage (about 30\%), whereas the average for the other architectures is only 4.2\%--8.2\%. This is because arm64 exhibits a richer set of relocation types and tightly coupled instruction patterns, such as \texttt{ADRP+ADD}~\cite{adrp}, which drive the decompiler into a linker-intensive regime that is particularly challenging for LLMs to handle.

\finding{Architecture does not merely change the rate of failure; it changes the \emph{failure mode}. On ARM64, $\sim$30\% of repair attempts shift into link-stage failure (vs.\ 4--8\% on x86/x64/ARM32), and the LLM repair agent cannot symbolize these ARM64-specific relocation idioms without engine-side support. \emph{Repair LLMs cannot rescue what the decompiler chose to keep as opaque address arithmetic.}}

% Architecture is not a mild nuisance factor in the Recompilability evaluation; it changes the failure mode itself. As shown in \cref{fig:compilability_by_arch}, arm64 is uniquely hard for GLM and Qwen, but not because it simply produces more compile failures. Instead, it pushes a large fraction of tasks into linker failure: 30.6\% for GLM and 30.3\% for Qwen, compared with only 4.2\%--8.2\% on the other three architectures. This indicates that many arm64 tasks can be brought through the compile phase but still fail to produce a final binary within the repair budget. By contrast, x64, x86, and arm32 all remain substantially easier for GLM and Qwen, with compile-success rates around 69\%--77\%. MiniMax is weaker on all four architectures, but arm64 is still its worst case, with only 18.5\% compile success and 61.7\% compile failure.

\begin{figure}[!ht]
  \centering
  \includegraphics[width=\linewidth]{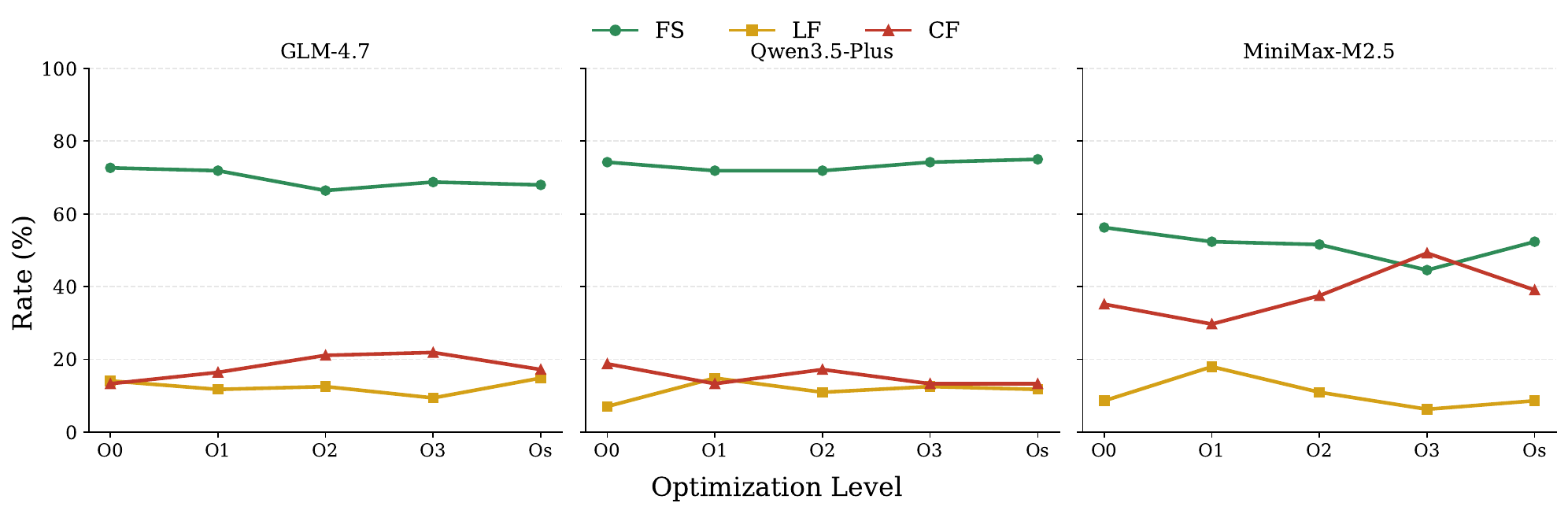}
  \caption{IDA-only FS/LF/CF rates across optimizations.}
  \label{fig:compilability_by_opt}
\end{figure}

\myparagraph{Optimization} Restricting the view to IDA also clarifies what optimization actually changes. In \cref{fig:compilability_by_opt}, GLM and Qwen remain relatively stable across \texttt{O0--Os}. 
MiniMax is much more sensitive, and the lost quality is mainly transferred to CF. Therefore, optimization will adjust the ratio of FS/LF/CF, but it cannot overturn the stronger effects dominated by architecture and code dimensions.

% MiniMax is much more sensitive. Its FS drops from 56.2\% at \texttt{O0} to 44.5\% at \texttt{O3}, and the lost mass goes primarily to CF, which rises from 35.2\% to 49.2\%, while LF remains secondary except for a spike to 18.0\% at \texttt{O1}. So even for the strongest decompiler, optimization is a secondary stressor: it modulates the FS/LF/CF mix, but it does not overturn the stronger architecture- and dimension-driven effects.

\myparagraph{Compiler or Debug} 
The changes caused by compilers and debugging symbols are even smaller. For GLM and Qwen, GCC is easier to fix than Clang. Binaries compiled using debugging symbols also show small but stable improvements. These effects do not change the ranking of decompilers, nor do they change the main sources of difficulty identified above.

\subsubsection{Multi-Factor Interactions}

\begin{figure}[!ht]
  \centering
  \includegraphics[width=\linewidth]{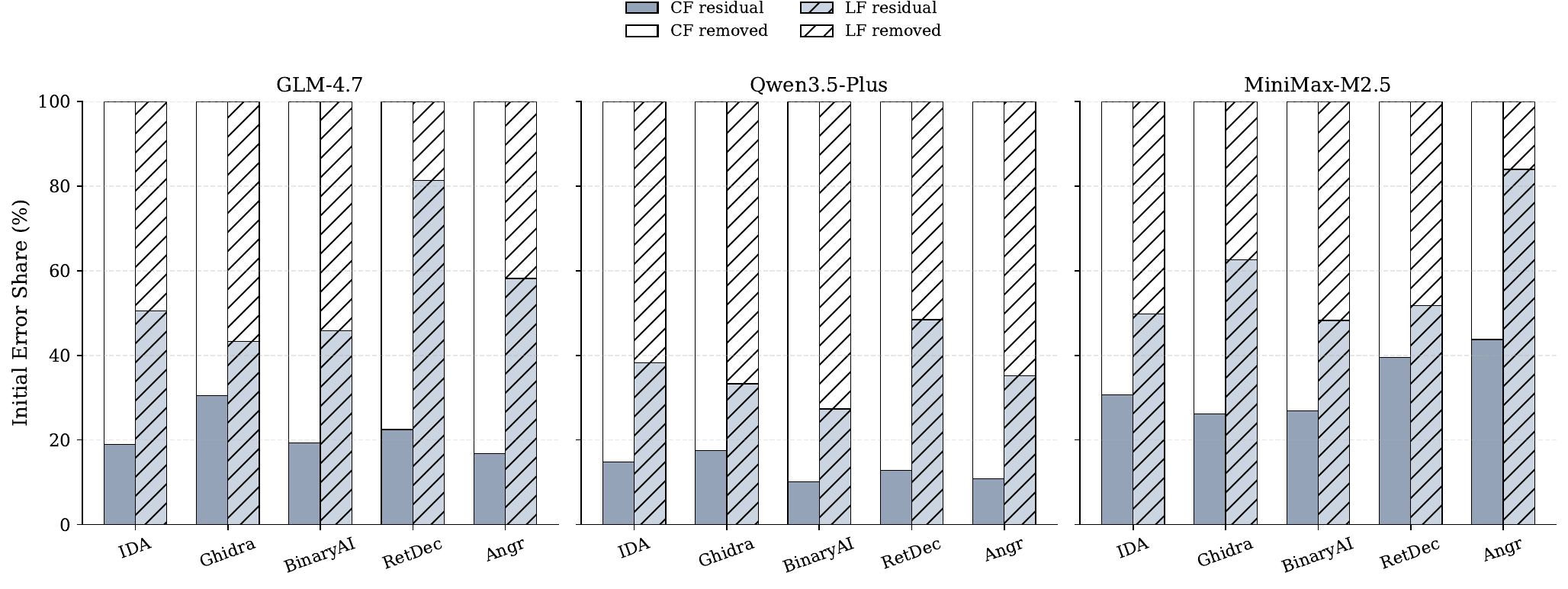}
  \caption{Effective repair effort inside failure cases. Each bar is normalized to the initial error total. The solid lower segment is the historical lowest residual error, and the hollow upper segment is the repaired error.}
  \label{fig:compilability_failure_effort_ratio}
\end{figure}

\myparagraphnew{In test cases of compilation and linking failures, is there also effective effort involved}
\cref{fig:compilability_failure_effort_ratio} shows that many failed tasks still eliminate a large share of their initial errors before the budget is exhausted. For compile-failed tasks, the drop from the hollow total bar to the solid historical-lowest segment is large across most decompilers: under Qwen, BinaryAI removes 89.9\% of initial compile errors in aggregate, Angr removes 89.2\%. Linker-failed tasks also frequently reflect near misses rather than dead ends. Under Qwen, BinaryAI removes 72.7\% of initial linker errors in aggregate and Ghidra removes 66.7\%.
% under GLM, Ghidra still removes 56.7\%. The main exceptions are weak convergence regimes such as MiniMax+angr at the linker stage, where only 16.0\% of linker errors are removed, and MiniMax+RetDec on compile-failed tasks, where the reduction drops to 60.5\%. 
The key point is that failure categories are not easy to define; many of these categories represent partial convergence, but still stop short of a final binary.

\subsection{Functionality}

We classify each (decompiler, binary) pair into one of four \textbf{program-level outcome categories} based on stdout-derived observations from the original and recompiled binaries:
\begin{itemize}[leftmargin=10pt,topsep=0pt,itemsep=0pt,parsep=0pt]
\item \textbf{Exact Stdout}: every observation parsed from the original binary's stdout has a matching observation in the recompiled binary's stdout, with identical values and ordering. Example: a Control-Flow case where the original prints \texttt{[CF-L1-03] result=42} and the recompiled binary prints the same line.
\item \textbf{Partial}: a strict non-empty subset of the original observations matches in the recompiled run, but at least one diverges. Example: a Memory-Operations case where the recompiled binary prints the correct results for the small-buffer subcases but produces different values for the aliasing subcase due to type-recovery error.
\item \textbf{Fail}: at least one observation contradicts (different value or wrong ordering) and no compensating evidence elsewhere supports equivalence; this includes the case where the recompiled binary crashes mid-run before reaching the comparison point. Example: an Object-Oriented C++ case where virtual dispatch is misrecovered and the recompiled binary segfaults on the first virtual call.
\item \textbf{Unsupported}: the case yields no comparable observable evidence (the original or recompiled binary produces no parsed observations, e.g., functions whose only side effect is silent state mutation that the case driver does not externalize). These are reported separately from \textbf{Fail} so that ``we measured divergence'' is not conflated with ``we lacked the evidence to measure.''
\end{itemize}
The function-level and instruction-level columns add finer evidence but do not change the program-level verdict.

\cref{tab:semantic_overview_glm} summarizes the functionality results of GLM-4.7. At the program level, only 1.2\% achieved perfect \texttt{Stdout} consistency, and 15.7\% maintained partial overlap.
IDA performed best, while RetDec had no perfect matches.
The function-level and instruction-level columns explain why program-level recovery is so difficult.
We found that only 61.9\% (1203) of the programs supported instrumented observations, with the rest allowing for crashes.
Of all the functions in these 1203 observable tasks, 78.5\% of the input/output signatures were aligned.
Instruction-level comparison based on sequence similarity was even more challenging, resulting in only 22\% similarity.

\begin{figure}[!ht]
  \centering
  \includegraphics[width=\linewidth]{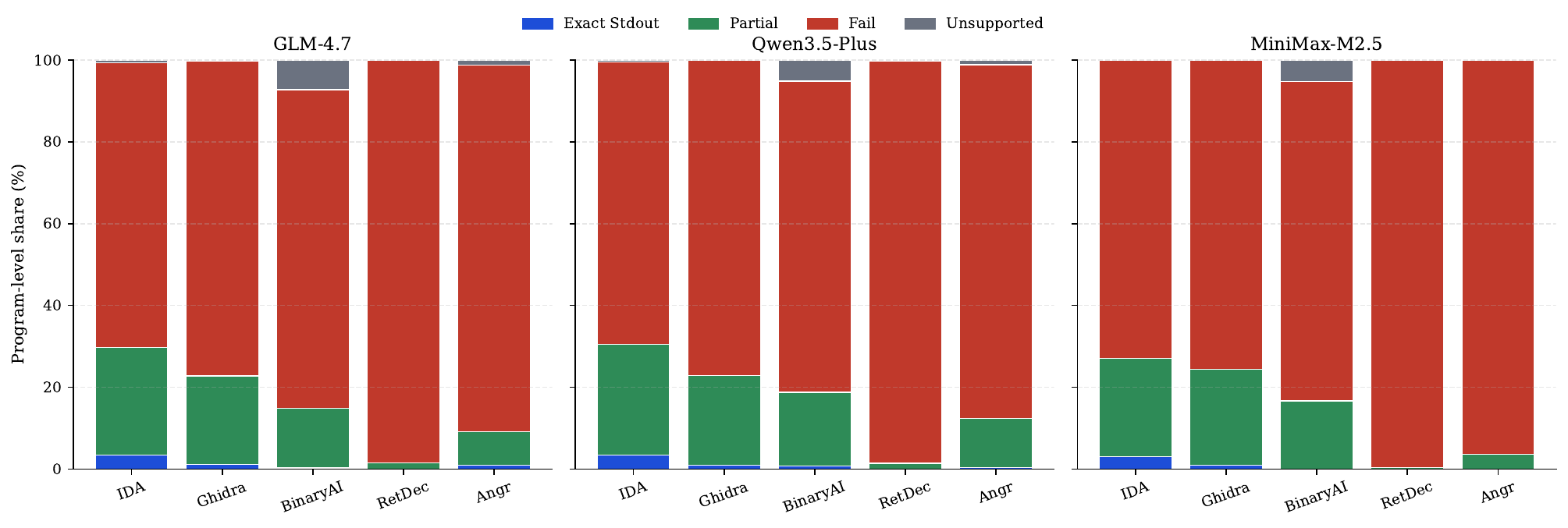}
  \caption{Program-level outcome distribution.}
  \label{fig:semantic_quality_by_decompiler}
\end{figure}

\cref{fig:semantic_quality_by_decompiler}  extends this procedural-level perspective to all three repair LLMs. For the LLMs, their performance is largely similar, indicating that decompilation quality is more important. IDA and Ghidra consistently maintain their leading position.

\finding{\emph{The reusability cliff is steep, reproducible, and not solvable by throwing more LLM at it.} The best decompiler--LLM pair (IDA + Qwen3.5-Plus) reaches URAF readability of 6.43/10 and 73.8\% recompilable binaries, but only 22.3\% of these reproduce \emph{any} program-level behavior and 1.2\% match \texttt{Stdout} exactly. The cliff between ``recompilable'' and ``behaviorally equivalent'' is $\sim$50 percentage points and is reproduced across three independent LLMs.}

\finding{\emph{The decompiler--LLM gap dwarfs the LLM--LLM gap by an order of magnitude.} At program-level functionality, switching decompiler under a fixed LLM moves Exact+Partial from 1.1\% (RetDec) to 22.3\% (IDA)---a 20$\times$ swing. Switching LLM under a fixed decompiler moves it from 13.9\% (IDA+MiniMax) to 22.3\% (IDA+Qwen)---a 1.6$\times$ swing. \emph{Investing in decompiler quality pays off more than investing in repair-LLM quality.}}

\begin{figure}[t]
  \centering
  \includegraphics[width=\linewidth]{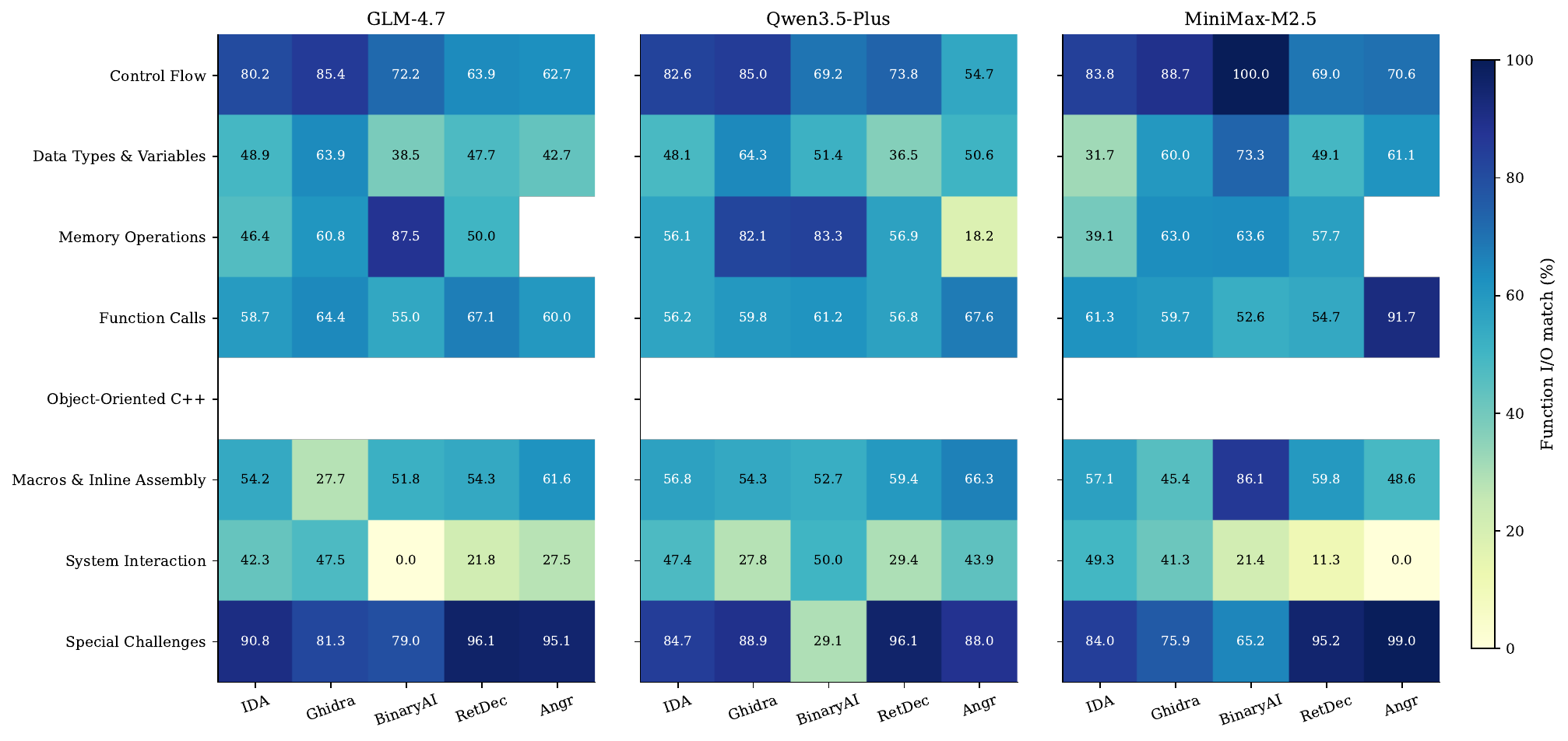}
  \caption{Function I/O match across dimensions. Empty cells indicate that
  there is no comparable matching call pair.}
  \label{fig:semantic_program_src_heatmap}
\end{figure}

\myparagraph{A Taxonomy of Semantic Failures} We manually inspected a stratified sample of 60 \textit{Fail} cases (12 per decompiler) to characterize \emph{why} program-level execution diverges. Four named failure modes account for the large majority of cases:
\circled{1} \textbf{Type-induced behavioral drift} (most common across all decompilers). Incorrect type recovery---most often signed-vs-unsigned confusion or pointer-vs-array degradation---causes silently wrong arithmetic or wrong control-flow decisions. The motivating example (\cref{fig:motivation}) is exactly this mode: Angr recovers all parameters as \texttt{unsigned int}, so the negative-input branches behave incorrectly even though the structure is preserved.
\circled{2} \textbf{Abstraction collapse.} Virtual-dispatch tables, multiple inheritance, and template instantiations are not reconstructed; the recompiled program either jumps to the wrong target or segfaults on the first virtual call. This mode dominates the \textit{Object-Oriented C++} dimension and explains the all-empty row of \cref{fig:semantic_program_src_heatmap}.
\circled{3} \textbf{Link-time symbol mismatch.} Compilation succeeds but the recompiled binary resolves a different symbol than the original (e.g., ARM64 \texttt{ADRP+ADD} relocations re-emitted as opaque address arithmetic that the linker cannot reconstruct), shifting failures into LF rather than CF and producing program-level divergence even when the function-level I/O check on surviving call pairs agrees.
\circled{4} \textbf{Side-effect erasure on void/no-return functions.} Functions whose only externally visible effect is a side effect (e.g., a logging call, a state mutation observed only via stdout) are scored as \texttt{Unsupported} when the case driver does not produce a comparable observation; this is a \emph{measurement gap}, not a decompiler defect, and we report \texttt{Unsupported} separately to keep the two distinct.
The first three modes correspond directly to the three actionable directions in RQ3: type-system reconstruction, language-aware front-ends for object-oriented ABIs, and IR-level symbolic relocation tracking.

% extends the program-level view to three LLMs. 

% Exact Stdout remains rare for every LLM and every decompiler, while the decompiler gap is far larger than the LLM gap. IDA and Ghidra consistently occupy the top tier, BinaryAI forms a middle tier, and Angr and especially RetDec are dominated by the \textit{Fail} bucket. After the rerun, \textit{Unsupported} remains low across the board, so most negative outcomes are evidence-backed mismatches rather than incomparable runs.

% \cref{tab:semantic_overview_glm} illustrates the overall effect of functional consistency on the GLM. 
% First, let's look at XX at the program level. 
% Then, at the function and instruction levels, XXX.
% \pz{decompiler}

\begin{table*}[!ht]
\centering
\caption{Functionality evaluation overview for GLM-4.7.}
\label{tab:semantic_overview_glm}
\resizebox{\textwidth}{!}{%
\begin{tabular}{lcccccccc}
\toprule
\multirow{2}{*}{\textbf{Decompiler}} & \multicolumn{4}{c}{\textbf{Program-level}} & \multicolumn{2}{c}{\textbf{Function-level}} & \multicolumn{2}{c}{\textbf{Instruction-level}} \\
\cmidrule(lr){2-5}\cmidrule(lr){6-7}\cmidrule(l){8-9}
& \textbf{Exact Stdout} & \textbf{Partial} & \textbf{Fail} & \textbf{Unsupported} & \textbf{Evidence Available} & \textbf{I/O Match} & \textbf{Evidence Available} & \textbf{Similarity} \\
\midrule
\textbf{IDA} 
& \cellcolor{gray!35}15/445 (3.4\%) 
& \cellcolor{gray!35}117/445 (26.3\%) 
& \cellcolor{gray!10}310/445 (69.7\%) 
& 3/445 (0.7\%) 
& 309/445 (69.4\%) 
& 83.3\% 
& \cellcolor{gray!35}222/445 (49.9\%) 
& \cellcolor{gray!35}28.9\% \\

\textbf{Ghidra} 
& 5/470 (1.1\%) 
& 102/470 (21.7\%) 
& 362/470 (77.0\%) 
& 1/470 (0.2\%) 
& \cellcolor{gray!35}327/470 (69.6\%) 
& \cellcolor{gray!10}69.1\% 
& 204/470 (43.4\%) 
& 24.6\% \\

\textbf{BinaryAI} 
& 1/385 (0.3\%) 
& 56/385 (14.5\%) 
& 300/385 (77.9\%) 
& \cellcolor{gray!35}28/385 (7.3\%) 
& \cellcolor{gray!10}183/385 (47.5\%) 
& 77.0\% 
& 122/385 (31.7\%) 
& 24.1\% \\

\textbf{RetDec} 
& \cellcolor{gray!10}0/339 (0.0\%) 
& \cellcolor{gray!10}5/339 (1.5\%) 
& \cellcolor{gray!35}334/339 (98.5\%) 
& \cellcolor{gray!10}0/339 (0.0\%) 
& 223/339 (65.8\%) 
& 74.8\% 
& 108/339 (31.9\%) 
& \cellcolor{gray!10}9.3\% \\

\textbf{Angr} 
& 3/306 (1.0\%) 
& 25/306 (8.2\%) 
& 274/306 (89.5\%) 
& 4/306 (1.3\%) 
& 161/306 (52.6\%) 
& \cellcolor{gray!35}91.3\% 
& \cellcolor{gray!10}93/306 (30.4\%) 
& 11.7\% \\
\midrule
\textbf{Total} & \textbf{24/1945 (1.2\%)} & \textbf{305/1945 (15.7\%)} & \textbf{1580/1945 (81.2\%)} & \textbf{36/1945 (1.9\%)} & \textbf{1203/1945 (61.9\%)} & \textbf{78.5\%} & \textbf{749/1945 (38.5\%)} & \textbf{22.0\%} \\
\bottomrule
\end{tabular}%
}
\end{table*}

% Regarding the LLMs, \cref{fig:semantic_quality_by_decompiler} shows that 
% \pz{LLMs}

\subsubsection{Single-Factor Effects}
We analyze the impact of each variable on functionality.

\myparagraph{Compiler} Compiler creates the clearest global separation---in the direction \emph{opposite} to readability. Clang substantially outperforms GCC at the observable program level, reaching 26.5\% Exact+Partial versus only 10.2\% for GCC. Clang also exposes more function evidence (69.2\% vs.\ 52.8\%) and much more instruction-level evidence (44.8\% vs.\ 29.7\%). Yet under URAF (\cref{tab:compiler_sensitivity}), GCC and Clang produce nearly identical readability scores (within 0.05 on every level). The same source compiled with two compilers produces decompiled output of comparable readability but $\sim$2.6$\times$ different probability of behavioral equivalence after repair---compiler choice is invisible to URAF but dominant for functionality.

\myparagraph{Dimension}
\cref{fig:semantic_program_src_heatmap} shows that source dimension is the strongest discriminator once we move from end-to-end stdout to function-level fidelity. \textit{Control Flow} and \textit{Special Challenges} frequently retain high matched-call I/O consistency across decompilers, with many cells above 80\%. The most important outlier is \textit{Object-Oriented C++}: every cell is empty because no comparable function I/O evidence survives at all, and its program-level outcome is correspondingly 100\% Fail. This demonstrates the difficulty of fixing C++ features, even with recompilation, which may not lead to correct execution.

\finding{Object-Oriented C++ is the \emph{unanimously} hardest dimension across all 5 decompilers, all 3 repair LLMs, and all 4 architectures: it is the only dimension that yields zero comparable function I/O evidence (an empty row in \cref{fig:semantic_program_src_heatmap}) \emph{and} the lowest URAF and recompilability scores. C++ ABI reconstruction (vtables, virtual dispatch, templates) requires language-aware front-end work---no amount of repair-LLM iteration substitutes for it.}

\myparagraph{Architecture} x64 is the most favorable architecture overall, with the highest Exact+Partial rate (19.3\%), the strongest Function I/O Match (84.7\%), and the highest instruction-level evidence availability (43.4\%). This is likely because LLMs are more familiar with x64 code, resulting in higher quality fixes.

\myparagraph{Optimization} Optimization level also exhibits a clear crossover, again \emph{against} the readability gradient. \texttt{O3}---the level with the \emph{lowest} URAF readability (\cref{fig:readability_by_opt})---gives the \emph{best} program-level recovery, with 24.8\% Exact+Partial and the lowest Fail rate (73.3\%). \texttt{O0}---the level with the highest readability---yields only 8.6\% Exact+Partial. Meanwhile \texttt{Os} collapses to only 35.8\% Function I/O Match, indicating that output-level survival and deep structural comparability can diverge sharply across optimizations.

\finding{Compilation choices that maximize readability are \emph{not} the ones that maximize behavioral preservation. Aggressive \texttt{O3} produces the lowest URAF readability but the highest program-level functionality (16.0\% vs.\ O0's 8.6\%); Clang produces lower readability than GCC but 2.6$\times$ higher functionality. \emph{Single-axis evaluations that proxy reusability by readability misrank both compiler and optimization choices.}}

\myparagraph{Debug} Debug symbols help observability more than they help every downstream metric uniformly. The \texttt{+g} setting modestly improves program-level recovery (19.1\% Exact+Partial vs.\ 15.9\% for no-g) and also improves instruction-level evidence and similarity.

\subsubsection{Multi-Factor Interactions} 
We analyze the reasons for semantic fidelity from the perspectives of program-level, function-level and instruction-level.

\myparagraphnew{What does program-level consistency mean} 
Program-level asks whether the repaired binary reproduces the observable benchmark behavior at the level of parsed test-case lines. The source-level breakdown shows that these mismatches come from at least two qualitatively different regimes. \textit{Object-Oriented C++} is a full abstraction-collapse regime. In contrast, \textit{Macros \& Inline Assembly} and \textit{Special Challenges} are also almost entirely Fail at the program level, yet they still preserve substantial downstream evidence. It suggests that many program-level failures are not caused by uniform semantic destruction; instead, the repaired program often retains islands of locally meaningful behavior while still breaking the externally visible contract through missing side effects, output-order changes, control-flow drift, or errors in code regions that never survive into the matched local slice.

\myparagraphnew{What does function-level consistency mean}
Function-level consistency answers a narrower and more conditional question: once comparable matched calls exist, do their local input/output signatures still agree? 
That pattern is the clearest evidence that function-level agreement is measuring a different phenomenon from end-to-end correctness. 
High Function I/O Match therefore means that the \emph{surviving comparable slice} of execution often remains locally plausible. It does \emph{not} mean that the repaired binary is globally correct.
\cref{fig:semantic_program_src_heatmap} makes this interpretation concrete. \textit{Special Challenges} reaches the highest aggregate Function I/O Match (83.6\%) while remaining almost entirely program-level Fail, which implies that its failures are often caused by global coordination problems rather than by uniformly wrong local computations. By contrast, \textit{System Interaction} and \textit{Macros \& Inline Assembly} stay weak even at the function level, which is consistent with semantics that depend more heavily on external effects, compiler-specific lowering, or low-level behavior that is harder to preserve under decompilation and repair. In short, the second layer is most useful for telling us \emph{how much coherent local behavior survives after global failure}, not for replacing the program-level verdict.

\myparagraphnew{What does instruction-level consistency mean}
Code changes inevitably lead to differences in instruction sequences; therefore, we use function call sequences for similarity analysis. Instruction-level consistency is the brake against over-interpreting high function-level scores. 
The clearest evidence again comes from the decompiler contrast.
Angr and RetDec still achieve 91.0\% and 74.0\% Function I/O Match on comparable call pairs, yet their program-level Exact+Partial rates are only 10.0\% and 1.1\%, and their instruction similarities are only 10.1\% and 10.9\%. 
 In contrast, IDA and Ghidra combine stronger program-level recovery capabilities with significantly higher instruction similarity.
  If high Function I/O Match alone meant robust semantic recovery, then Angr and RetDec should have looked much better at the program level than they actually do. They do not. The more consistent interpretation is that those decompilers often preserve enough recognizable call boundaries to create a small set of locally comparable traces, but not enough structural fidelity to sustain correct whole-program behavior.

\begin{table}[!t]
\centering
\caption{Efficiency summary.}
\label{tab:efficiency_summary}
\resizebox{\columnwidth}{!}{%
\begin{tabular}{llrrrrrr}
\hline
\textbf{LLM} & \textbf{Stage} & \textbf{Token (M)} & \textbf{Avg. Token (K)} & \textbf{Med. Token (K)}  & \textbf{Avg. Time (s)} & \textbf{Med Time (s)} \\
\hline
\multirow{2}{*}{GLM-4.7}
  & Readability    & 92.1  & 30.5  & 29.1  & 76.8 & 54.9 \\
  & Recompilability  & 1{,}377.8  & 456.1 & 301.6 & 1473.5 & 871.0 \\
  & Functionality  &   - & -  & -  & 12.3 & 21.0 \\
\hline
\multirow{2}{*}{Qwen3.5-Plus}
  & Readability    & 98.0  & 32.4  & 29.6  & 62.4 & 52.9 \\
  & Recompilability  & 1{,}453.7  & 481.0 & 266.9  & 826.0 & 430.1 \\
    & Functionality  &   - & -  & -  & 15.3 & 29.0 \\
\hline
\multirow{2}{*}{MiniMax-M2.5}
  & Readability    & 82.5  & 27.3  & 26.0  & 59.4 & 55.6 \\
  & Recompilability  & 1{,}683.0  & 557.8 & 500.1  & 1046.9 & 1108.9 \\
    & Functionality  &   - & -  & -  & 14.1 & 24.6 \\
\hline
\end{tabular}%
}
\end{table}

\subsection{Efficiency Analysis}

The efficiency view in this subsection is reported for transparency and reproducibility, not as a benchmark of the LLMs themselves: the object of study is the decompiler, and the LLMs are instruments. We therefore highlight per-decompiler costs and treat per-LLM differences as inter-rater spread.

Efficiency evaluation considers both token consumption and wall-clock time, as summarized in \cref{tab:efficiency_summary}. 
Recompilability repair dominates the overhead. To prevent non-terminating repair loops, we cap the maximum number of repair iterations at 50. It is worth noting that efficiency is influenced by multiple factors, which are not the primary focus of this paper and are therefore not explored in depth. Intuitively, the overhead is closely tied to the quality and conciseness of the decompiler output: shorter and higher-quality outputs generally lead to lower repair cost. 
Consistent with this observation, IDA incurs the lowest overhead, whereas Angr exhibits  higher cost due to less structured outputs. In contrast, RetDec often fails to participate effectively in the evaluation, as its excessively long outputs frequently exceed the LLM context window.

\newcommand{\myreason}[1]{
\noindent\uline{#1:}\quad
}

\section{Research Questions}
Guided by the evaluation in \cref{secfinding} and the boxed findings highlighted there, we organize the discussion around four sharpened research questions. Each is answered with concrete numbers from the same 240-function $\times$ 640-binary $\times$ 5-decompiler $\times$ 3-LLM matrix, so the answers are directly comparable.

\myparagraphnew{RQ1: How wide is the reusability gap between ``readable'' and ``behaviorally equivalent'' decompiled code, and is the gap reproducible}

\myreason{Answer}
The gap is wide, monotone, and reproducible across three independent repair LLMs. On the same benchmark, the best decompiler (IDA) is scored 6.43/10 on URAF readability, repairs to a linkable binary in 65--74\% of cases depending on the repair LLM, but reproduces \emph{any} program-level behavior in only 20.6--22.3\% of those cases and matches the original \texttt{Stdout} exactly in only 1.2\%. The cliff between \textit{recompilable} and \textit{behaviorally equivalent} is roughly 50 percentage points, and the per-decompiler rank order is preserved under all three judges (head-to-tail ordering---IDA top, Angr/RetDec bottom---is unanimous on all five readability levels; mean cross-judge Spearman $\rho = 0.88$). The implication is that ``moderate readability'' overstates downstream utility by roughly $3\times$: a benchmark that stops at readability would license tools that, in practice, almost never produce a behaviorally equivalent rebuilt binary.

\myreason{Implication for decompiler developers}
Maintain readability as a necessary but \emph{insufficient} milestone. Treat the cliff between recompilability and program-level functionality as the central engineering target: surface-clean code that does not behave like the original is not a partial success but a complete one for the wrong objective.

\myparagraphnew{RQ2: Do compilation choices that improve readability also improve behavioral fidelity, or do the metrics decouple}

\myreason{Answer}
They sharply decouple, in two directions that prior single-axis evaluations would have misranked. (i) Compiler frontend: GCC and Clang produce nearly identical URAF readability (within 0.05 on every level), yet Clang reaches 26.5\% Exact+Partial program-level functionality vs.\ GCC's 10.2\%---a $2.6\times$ gap that URAF cannot see. (ii) Optimization level: \texttt{O3} produces the \emph{lowest} URAF readability (5.32) but the \emph{highest} functionality (16.0\% Exact+Partial); \texttt{O0} produces the highest readability but only 8.6\% functionality. Architecture and debug symbols move all three metrics in the same direction, but compiler and optimization invert. This means that the standard practice of using readability as a proxy for downstream usefulness is not just imprecise---it is \emph{wrong} on exactly the two compilation knobs that practitioners control most often.

\myreason{Implication for benchmark designers}
Evaluations that report only readability or only recompilability cannot rank tools or compilation pipelines correctly when the downstream task is behavioral reuse. Multi-axis reporting (or, at minimum, a behavioral verification stage) is necessary.

\myparagraphnew{RQ3: Where should engineering investment go---into better decompilers or better repair LLMs}

\myreason{Answer}
Into decompilers. At program-level functionality, switching decompiler under a fixed LLM moves Exact+Partial from 1.1\% (RetDec+Qwen) to 22.3\% (IDA+Qwen)---a $20\times$ swing. Switching LLM under a fixed decompiler moves the same number from 13.9\% (IDA+MiniMax) to 22.3\% (IDA+Qwen)---a $1.6\times$ swing. The per-decompiler rank order is preserved across all three LLMs. Cost data points the same way: MiniMax-M2.5 consumes 22\% more tokens than GLM-4.7 but produces 41\% fewer successful binaries---a weaker repair LLM does not stall on solvable instances, it burns budget oscillating on the unsolvable ones. \emph{The bottleneck is the decompiled input, not the repair LLM.}

\myreason{Implication for the LLM-decompilation community}
Function-level natural-style decoding (LLM4Decompile, DecLLM) addresses the small slice of variance attributable to the repair LLM. The much larger slice attributable to the decompiler engine---type recovery, ABI handling, relocation symbolization---requires engine-level rather than prompt-level innovation.

\myparagraphnew{RQ4: When the pipeline fails, where in it does the failure originate, and which failures can post-hoc LLM repair compensate for}

\myreason{Answer}
We mined 428k error instances across all repair traces and inspected a stratified sample of 60 failure cases. Three sources, with very different reparability:
\circled{1} \textbf{Syntactic noise} (Syntax Error 31.0\%, Undeclared Identifier 7.6\%, Implicit Function Declaration 2.8\%, etc.). This is what LLM repair is good at: surface-level patching that converts raw decompiler output into something a compiler accepts. Most successful repairs are concentrated here.
\circled{2} \textbf{Type-system collapse} (Conflicting Types, Incompatible Pointer Type, Unknown Type, Incomplete Type, Member Access Error, Void Value Error, collectively $\sim$19\%). This is partially repairable but with diminishing returns: even when the LLM patches a type to make the code compile, the resulting types are often incorrect in ways that downstream behavior exposes---the \textit{Type-induced behavioral drift} failure mode dominates our 60-case sample and is the single largest non-syntactic cost.
\circled{3} \textbf{Irreversible upstream losses} (ARM64-specific relocation patterns; C++ vtable/dispatch reconstruction; symbol-level external interface). LLM repair is structurally unable to fix these: it cannot invent the symbol the linker is missing, it cannot re-derive a vtable that the decompiler never emitted, and it cannot symbolize an \texttt{ADRP+ADD} pair that was left as opaque address arithmetic. These categories explain why Object-Oriented C++ collapses to zero comparable function I/O evidence and why ARM64 shifts $\sim$30\% of tasks into the link-stage failure regime.

\myreason{Implication}
The three priority directions for decompiler-side investment, in descending impact, are: (a) composite-type and structural type recovery that survives symbol stripping (largest residual cost in our data); (b) IR-level symbolic relocation tracking that does not collapse \texttt{ADRP+ADD} idioms into opaque arithmetic (responsible for the ARM64 failure-mode shift); (c) C++ ABI-aware front-end passes for vtable and dispatch reconstruction (responsible for the unanimous OOP collapse). Each is a categorical capability the repair LLM cannot provide.

\section{Threats to Validity}\label{secthreats}

\myparagraph{Construct validity: LLM-as-judge readability scoring}
URAF assigns readability via three independent LLM judges. A single judge could in principle bias the absolute scale or favor a particular code style. We mitigate this in three ways: (i) we use three judges (GLM-4.7, Qwen3.5-Plus, MiniMax-M2.5) and report cross-judge spread; (ii) the judges are anchored to a five-band qualitative rubric (\cref{tab:readiablity}) over 18 explicit sub-dimensions, so any judge that drifts can be cross-checked at the sub-dimension level; (iii) URAF rankings are triangulated against the deterministic recompilability and functionality pipelines, which do not use an LLM as the oracle (the compiler and runtime are the oracles). The cross-pipeline rank coupling reported in \cref{secfinding} is the strongest available indirect evidence that URAF is measuring a real signal rather than judge artifact. We do not claim absolute calibration of URAF scores against human raters; the framework is designed for \emph{relative} comparison across decompilers and configurations. A human-expert calibration study, in which reverse engineers score a stratified sample under the same 18-sub-dimension rubric and inter-rater agreement is computed against the LLM judges, would yield an absolute anchor for URAF and is a natural next step that is bounded by human-study cost rather than by the framework.

\myparagraph{Internal validity: coverage of dynamic verification}
Differential dynamic tracing is observational and is bounded by the paths exercised by each case's driver. Branches the driver does not enter are silently \emph{not} disconfirmed by our metric. We address this in four ways: (i) every benchmark case is paired with a driver designed to exercise the function under test directly rather than via incidental paths, and we verified empirically (\cref{secfinding}, dynamic verification section) that the per-dimension driver coverage is 95.5\%--100\% of expected functions, so the residual measurement gap is small in absolute terms; (ii) for void/no-return-value functions we use side-effect channels (hooked \texttt{write} on \texttt{fd=1} associated with the call frame) rather than awarding spurious matches; (iii) we report function-level and instruction-level agreement as conditional on observability and, when those signals diverge from program-level outcomes (e.g., Angr's high local I/O match coexisting with low program-level recovery), we explicitly interpret the divergence rather than collapse it into a single number; (iv) coverage alone bounds what \emph{could} be observed; a complementary question is what is \emph{reliably detected once observed}. To answer this in future work, we plan a mutation-testing study with a pre-registered protocol: starting from each of the 240 atomic source functions, generate $N=10$ mutants per function (2,400 mutants total) using four decompilation-relevant operator families---(a) \textit{type drift} (signed$\leftrightarrow$unsigned, narrow$\leftrightarrow$wide integer, pointer$\leftrightarrow$array of same element type), (b) \textit{control-flow flip} (branch condition negation, loop bound off-by-one), (c) \textit{ABI/calling-convention} (argument-order swap, return-register class change), and (d) \textit{side-effect omission} (drop the last \texttt{printf}/\texttt{write} on a success path). Each mutant is compiled and run through the same Frida pipeline as the original; the metric is detection rate, defined as the fraction of mutants for which the program-level outcome moves from Exact Stdout to Partial/Fail. Operators (a)--(c) target the failure modes our 60-case manual study (\cref{secfinding}) found dominant; operator (d) calibrates the void/side-effect channel specifically. Reporting detection rate per operator family will quantify the discriminative power of the pipeline against an injected-fault oracle, separately from driver coverage. This is a substantial follow-on study; we report the protocol here so that the design is auditable and the result is comparable across future replications.

\myparagraph{Internal validity: repair budget and cost metric}
The 50-iteration repair cap could in principle cut off runs that would have eventually succeeded. Pilot data show that cumulative success flattens after roughly 20--30 iterations and the long tail beyond 30 typically reflects oscillating or stuck repair patterns rather than slow progress, so the cap is in the post-plateau regime; we report the three-tier outcome (FS/LF/CF) precisely so that the headline metric is invariant to small changes in the cap. We also report the normalized fraction of initial errors removed (\cref{fig:compilability_failure_effort_ratio}) to show that ``failure'' is not all-or-nothing.

\myparagraph{External validity: benchmark scope}
\proj is a 240-function atomic suite compiled into 640 binaries across four architectures, two compilers, five optimization levels, and two debug settings. It is deliberately fine-grained to support precise localization of failure modes, and its scope is mainstream compilation pipelines without obfuscation, packing, or adversarial transformations. Generalization to wild stripped or obfuscated binaries (control-flow flattening, opaque predicates, virtualized obfuscation) is an explicit out-of-scope item; the framework is extensible to those settings via the same compilation matrix and dynamic-verification pipeline (see \cref{subsecbench}). Our recompilability and functionality numbers should therefore be read as \emph{best-case} bounds for current decompilers under standard compilation, not as ceilings under adversarial conditions.

\myparagraph{External validity: scope of decompiler comparison}
The five decompilers in the program-level evaluation (IDA, Ghidra, RetDec, Angr, BinaryAI) are whole-program traditional decompilers and span both proprietary (IDA, BinaryAI) and open-source (Ghidra, RetDec, Angr) systems, so the conclusions are not tied to a single licensing model. We do not include the Radare2 family (\texttt{r2dec}, \texttt{r2ghidra}) separately because its decompilation core is a thin wrapper over either a custom IR backend or Ghidra's engine, and it would not constitute an additional independent point in our comparison; the underlying engines we evaluate already cover that capability surface. LLM-based, function-granularity decompilers (LLM4Decompile~\cite{tan2024llm4decompile}, DecLLM~\cite{wong2025decllm}) are evaluated on URAF and function-level functionality but cannot be evaluated on program-level recompilability without additional harnessing, because they do not produce a linkable program. We keep the two unit bases separate to avoid mixing comparable and non-comparable measurements. As LLM-based decompilers mature toward whole-program output, extending the program-level metrics to them is a natural follow-up.

\section{Discussion}

\myparagraph{Obfuscation and adversarial settings}
Our study targets mainstream compilation pipelines and does not cover code obfuscation or adversarial reverse-engineering settings (e.g., control-flow flattening, opaque predicates). Such defenses deliberately destroy structural and semantic regularities relied on by decompilers, fundamentally changing the problem and raising distinct questions (e.g., robustness to targeted transformations, attack–defense dynamics). Systematic evaluation under obfuscation is therefore beyond our current scope. We position our benchmark as a foundation for studying decompilers under standard compilation settings and leave obfuscation-focused to future work.

\myparagraph{Data leakage}
 For LLM-based tools, a key concern is data leakage: future models may train on our benchmark or related code, artificially inflating absolute performance. This risk is inherent to any public benchmark~\cite{guan2025benchmarkstillusefuldynamic}. We chiefly rely on relative comparisons across decompilers and experimental conditions, which are less sensitive to uniform score inflation from training-data contamination. Our goal is to characterize and compare current decompilers and identify directions for improvement, not to claim permanent absolute performance bounds. In the longer term, mitigating data leakage will require efforts from model providers, including continuous benchmark renewal and explicit training-data hygiene to limit overlap between evaluation datasets and training corpora.
 
% When evaluating LLM-based tools, a natural concern is potential data leakage: future models may be trained on our benchmark or closely related code, thereby artificially inflating their absolute performance. In principle, this risk cannot be fully avoided for any publicly available benchmark. In our setting, however, we primarily rely on relative comparisons across decompilers and experimental conditions, which are less sensitive to uniform score inflation caused by training-data contamination. Moreover, the main contribution of our analysis is to characterize and compare the capabilities of current decompilers and to highlight concrete directions for their evolution, rather than to claim absolute performance bounds that must hold indefinitely. In the longer term, mitigating data leakage for LLM-based tools will require coordinated efforts from model providers, including continuous benchmark renewal and explicit training-data hygiene to limit overlap between evaluation datasets and model training corpora.

\myparagraph{Benchmark coverage and extensibility} 
Our benchmark emphasizes factors that directly affect code structure, such as optimization level, debug symbols, and diverse source-code characteristics. Other aspects, including additional instruction set architectures (e.g., MIPS, PowerPC) and binary formats (e.g., PE), also challenge decompiler generality but are less central to our core concerns of readability, recompilability, and functionality. The benchmark is designed to be scalable: by adapting the compilation scripts, it can be extended to cover further architectures and formats.

% Our current benchmark primarily targets factors that directly affect code structure, such as optimization levels, debug symbol, and diverse source-code dimensions. 
% Other factors, such as additional instruction set architectures (e.g., MIPS, PowerPC) and binary formats (e.g., PE on Windows), also pose challenges to the generality of decompilers. 
% However, these are not critical to our core concerns such as readability, recompilability, and functionality. 
% Importantly, the benchmark is designed to be scalable and can be modified to cover these architecture and format-related aspects by modifying the compilation scripts.

\section{Related Work}

\myparagraph{Decompiler Technique}
Binary decompilation aims to recover high-level, human-readable source code from machine code. Early work addressed key structural problems such as control-flow reconstruction~\cite{cifuentes1994reverse,brumley2013native} and data type recovery~\cite{schwartz2011q}, 
later augmented with advanced static analysis and type inference. A major step was the adoption of IRs, enabling architecture-agnostic analysis; prominent tools such as IDA Pro~\cite{ida} and Ghidra~\cite{ghidra} build on this approach. 
To enrich the recovered semantics, researchers have integrated taint analysis for data-flow tracking~\cite{song2008bitblaze}, pattern matching for library identification~\cite{gao2021lightweight,yakdan2015no},  and, more recently, machine learning for idiomatic pattern recognition~\cite{katz2018using,chen2022augmenting,lacomis2019dire,zhu2024tygr,sha2025llasm}. However, even when functionally correct, traditional decompilers often fail to match the abstraction and style of human-written code, creating a readability gap. This has motivated deep-learning and LLM-based methods that improve code aesthetics by predicting meaningful names and refactoring structure~\cite{liu2025function,he2018debin,xie2024resym,hu2024degpt,jin2022symlm}. LLMs further extend capabilities by automatically repairing and refactoring decompiled code to ensure recompilability and executability~\cite{wong2025decllm,tan2024llm4decompile}, substantially increasing its usefulness for downstream tasks.

\myparagraph{Decompiler Evaluation}
As decompilation techniques evolved, so did evaluation methodologies. Early work mainly examined semantic correctness, identifying defects that cause behavioral discrepancies between original and decompiled code~\cite{liu2020far,dramko2024taxonomy}. Subsequent studies evaluated additional quality attributes: Soni et al.~\cite{10.1145/3733822.3764675} measured type inference accuracy, and Cao et al.~\cite{10.1145/3520312.3534867} analyzed how compiler optimizations affect readability. More comprehensive benchmarks such as DecompileBench~\cite{gao2025decompilebench} combine fuzz testing to validate semantic correctness and already recognize LLMs as part of the decompilation pipeline. A parallel line of work targets LLM-based decompilers directly, including LLM4Decompile~\cite{tan2024llm4decompile} and DecLLM~\cite{wong2025decllm}, which operate at function granularity and emphasize natural-style code generation rather than program-level reconstruction. These efforts, however, typically focus on individual aspects (e.g., correctness, types, or readability), and the program-level versus function-level evaluation gap between traditional and LLM-based decompilers has not been bridged in a single framework. \proj fills this gap: we evaluate whole-program traditional decompilers across readability, recompilability, and functionality at program level, and extend URAF and function-level functionality to LLM-based decompilers on their natural unit, so the two classes of tools are compared on commensurable measurements rather than ignored or conflated.

\section{Conclusion}

We propose a reusability-driven paradigm for binary decompiler evaluation and instantiate it in \proj, the first framework that jointly evaluates readability, recompilability, and functionality on the same benchmark. On a 240-function $\times$ 640-binary $\times$ 5-decompiler $\times$ 3-LLM matrix, \proj surfaces four contributive findings that single-axis evaluations miss: a $\sim$50-percentage-point cliff between recompilability and behavioral equivalence; a counter-intuitive decoupling of readability and functionality under compiler and optimization choices ($2.6\times$ functionality gap between Clang and GCC at near-identical URAF readability; \texttt{O3} yields lowest readability but highest functionality); a $20\times$ cross-decompiler swing in functionality versus only a $1.6\times$ cross-LLM swing, indicating where engineering investment should go; and a three-category failure decomposition that isolates which decompiler defects are repairable by post-hoc LLM and which are not. These results reframe decompiler quality as a multi-axis continuum, give practitioners scenario-based tool-selection guidance, and identify three concrete engineering priorities---composite-type reconstruction, IR-level symbolic relocation, and C++ ABI-aware front-end passes---that move binary decompilation from empirical craft toward an engineering science.

\section{Data-Availability Statement}
The data and codes used in this study are publicly available at: 
\url{https://github.com/codefuse-ai/CodeFuse-DeBench}.

\bibliographystyle{colm2024_conference}
\bibliography{refer}

@inproceedings{10.1145/3520312.3534867,
author = {Armengol-Estap\'{e}, Jordi and Woodruff, Jackson and Brauckmann, Alexander and Magalh\~{a}es, Jos\'{e} Wesley de Souza and O'Boyle, Michael F. P.},
title = {ExeBench: an ML-scale dataset of executable C functions},
year = {2022},
isbn = {9781450392730},
publisher = {Association for Computing Machinery},
address = {New York, NY, USA},
url = {https://doi.org/10.1145/3520312.3534867},
doi = {10.1145/3520312.3534867},
booktitle = {Proceedings of the 6th ACM SIGPLAN International Symposium on Machine Programming},
pages = {50–59},
numpages = {10},
location = {San Diego, CA, USA},
series = {MAPS 2022}
}

@inproceedings{liu2022finding,
  title={Finding vulnerabilities in internal-binary of firmware with clues},
  author={Liu, Puzhuo and Fang, Dongliang and Qin, Chuan and Cheng, Kai and Lv, Shichao and Zhu, Hongsong and Sun, Limin},
  booktitle={ICC 2022-IEEE International Conference on Communications},
  pages={5397--5402},
  year={2022},
  organization={IEEE}
}

@INPROCEEDINGS {bridge2026,
author = { Peng, Jiaqian and Liu, Puzhuo and Zeng, Yicheng and Cheng, Kai and Liu, Yongji and Yang, Yun and Zhu, Hongsong },
booktitle = { 2026 IEEE Symposium on Security and Privacy (SP) },
title = {{ Bridge: High-Order Taint Vulnerabilities Detection in Linux-based IoT Firmware }},
year = {2026},
volume = {},
ISSN = {2375-1207},
pages = {2659-2678},
doi = {10.1109/SP63933.2026.00001},
url = {https://doi.ieeecomputersociety.org/10.1109/SP63933.2026.00001},
publisher = {IEEE Computer Society},
address = {Los Alamitos, CA, USA},
month =May}

@inproceedings{gao2025decompilebench,
  title={DecompileBench: A comprehensive benchmark for evaluating decompilers in real-world scenarios},
  author={Gao, Zeyu and Cui, Yuxin and Wang, Hao and Qin, Siliang and Wang, Yuanda and Bolun, Zhang and Zhang, Chao},
  booktitle={Findings of the Association for Computational Linguistics: ACL 2025},
  pages={23250--23267},
  year={2025}
}

@misc{tan2025decompilebench,
      title={Decompile-Bench: Million-Scale Binary-Source Function Pairs for Real-World Binary Decompilation}, 
      author={Hanzhuo Tan and Xiaolong Tian and Hanrui Qi and Jiaming Liu and Zuchen Gao and Siyi Wang and Qi Luo and Jing Li and Yuqun Zhang},
      year={2025},
      eprint={2505.12668},
      archivePrefix={arXiv},
      primaryClass={cs.SE},
      url={https://arxiv.org/abs/2505.12668}, 
}

@inproceedings{hu2024degpt,
  title={Degpt: Optimizing decompiler output with llm},
  author={Hu, Peiwei and Liang, Ruigang and Chen, Kai},
  booktitle={Proceedings 2024 Network and Distributed System Security Symposium},
  volume={267622140},
  year={2024}
}

@inproceedings{10.1145/3733822.3764675,
author = {Soni, Vedant and Dutcher, Audrey and Bao, Tiffany and Wang, Ruoyu},
title = {Benchmarking Binary Type Inference Techniques in Decompilers},
year = {2025},
isbn = {9798400719103},
publisher = {Association for Computing Machinery},
address = {New York, NY, USA},
url = {https://doi.org/10.1145/3733822.3764675},
doi = {10.1145/3733822.3764675},
booktitle = {Proceedings of the 2025 Workshop on Software Understanding and Reverse Engineering},
pages = {48–60},
numpages = {13},
location = {
},
series = {SURE '25}
}

@article{liu2025llm,
  title={Llm-powered static binary taint analysis},
  author={Liu, Puzhuo and Sun, Chengnian and Zheng, Yaowen and Feng, Xuan and Qin, Chuan and Wang, Yuncheng and Xu, Zhenyang and Li, Zhi and Di, Peng and Jiang, Yu and others},
  journal={ACM Transactions on Software Engineering and Methodology},
  volume={34},
  number={3},
  pages={1--36},
  year={2025},
  publisher={ACM New York, NY}
}

@inproceedings{brumley2013native,
  title={Native x86 decompilation using $\{$Semantics-Preserving$\}$ structural analysis and iterative $\{$Control-Flow$\}$ structuring},
  author={Brumley, David and Lee, JongHyup and Schwartz, Edward J and Woo, Maverick},
  booktitle={22nd USENIX Security Symposium (USENIX Security 13)},
  pages={353--368},
  year={2013}
}

@inproceedings{liu2020far,
  title={How far we have come: Testing decompilation correctness of C decompilers},
  author={Liu, Zhibo and Wang, Shuai},
  booktitle={Proceedings of the 29th ACM SIGSOFT International Symposium on Software Testing and Analysis},
  pages={475--487},
  year={2020}
}

@inproceedings{yakdan2015no,
  title={No More Gotos: Decompilation Using Pattern-Independent Control-Flow Structuring and Semantic-Preserving Transformations.},
  author={Yakdan, Khaled and Eschweiler, Sebastian and Gerhards-Padilla, Elmar and Smith, Matthew},
  booktitle={NDSS},
  year={2015}
}

@inproceedings{10.1145/3650212.3652144,
author = {Cao, Ying and Zhang, Runze and Liang, Ruigang and Chen, Kai},
title = {Evaluating the Effectiveness of Decompilers},
year = {2024},
isbn = {9798400706127},
publisher = {Association for Computing Machinery},
address = {New York, NY, USA},
url = {https://doi.org/10.1145/3650212.3652144},
doi = {10.1145/3650212.3652144},
booktitle = {Proceedings of the 33rd ACM SIGSOFT International Symposium on Software Testing and Analysis},
pages = {491–502},
numpages = {12},
location = {Vienna, Austria},
series = {ISSTA 2024}
}

@inproceedings{dramko2024taxonomy,
  title={A taxonomy of C decompiler fidelity issues},
  author={Dramko, Luke and Lacomis, Jeremy and Schwartz, Edward J and Vasilescu, Bogdan and Le Goues, Claire},
  booktitle={33rd USENIX Security Symposium (USENIX Security 24)},
  pages={379--396},
  year={2024}
}

@article{sha2025llasm,
  title={llasm: Naming functions in binaries by fusing encoder-only and decoder-only llms},
  author={Sha, Zihan and Wang, Hao and Gao, Zeyu and Shu, Hui and Zhang, Bolun and Wang, Ziqing and Zhang, Chao},
  journal={ACM Transactions on Software Engineering and Methodology},
  volume={34},
  number={4},
  pages={1--22},
  year={2025},
  publisher={ACM New York, NY}
}

@inproceedings{zhu2024tygr,
  title={$\{$TYGR$\}$: Type Inference on Stripped Binaries using Graph Neural Networks},
  author={Zhu, Chang and Li, Ziyang and Xue, Anton and Bajaj, Ati Priya and Gibbs, Wil and Liu, Yibo and Alur, Rajeev and Bao, Tiffany and Dai, Hanjun and Doup{\'e}, Adam and others},
  booktitle={33rd USENIX Security Symposium (USENIX Security 24)},
  pages={4283--4300},
  year={2024}
}

@inproceedings{gao2021lightweight,
  title={A lightweight framework for function name reassignment based on large-scale stripped binaries},
  author={Gao, Han and Cheng, Shaoyin and Xue, Yinxing and Zhang, Weiming},
  booktitle={Proceedings of the 30th ACM SIGSOFT International Symposium on Software Testing and Analysis},
  pages={607--619},
  year={2021}
}

@inproceedings{lacomis2019dire,
  title={Dire: A neural approach to decompiled identifier naming},
  author={Lacomis, Jeremy and Yin, Pengcheng and Schwartz, Edward and Allamanis, Miltiadis and Le Goues, Claire and Neubig, Graham and Vasilescu, Bogdan},
  booktitle={2019 34th IEEE/ACM International Conference on Automated Software Engineering (ASE)},
  pages={628--639},
  year={2019},
  organization={IEEE}
}

@inproceedings{chen2022augmenting,
  title={Augmenting decompiler output with learned variable names and types},
  author={Chen, Qibin and Lacomis, Jeremy and Schwartz, Edward J and Le Goues, Claire and Neubig, Graham and Vasilescu, Bogdan},
  booktitle={31st USENIX Security Symposium (USENIX Security 22)},
  pages={4327--4343},
  year={2022}
}

@inproceedings{jin2022symlm,
  title={Symlm: Predicting function names in stripped binaries via context-sensitive execution-aware code embeddings},
  author={Jin, Xin and Pei, Kexin and Won, Jun Yeon and Lin, Zhiqiang},
  booktitle={Proceedings of the 2022 ACM SIGSAC Conference on Computer and Communications Security},
  pages={1631--1645},
  year={2022}
}

@inproceedings{liu2023fits,
  title={Fits: Inferring intermediate taint sources for effective vulnerability analysis of iot device firmware},
  author={Liu, Puzhuo and Zheng, Yaowen and Sun, Chengnian and Qin, Chuan and Fang, Dongliang and Liu, Mingdong and Sun, Limin},
  booktitle={Proceedings of the 28th ACM International Conference on Architectural Support for Programming Languages and Operating Systems, Volume 4},
  pages={138--152},
  year={2023}
}

@inproceedings{wu2024your,
  title={Your firmware has arrived: A study of firmware update vulnerabilities},
  author={Wu, Yuhao and Wang, Jinwen and Wang, Yujie and Zhai, Shixuan and Li, Zihan and He, Yi and Sun, Kun and Li, Qi and Zhang, Ning},
  booktitle={33rd USENIX Security Symposium (USENIX Security 24)},
  pages={5627--5644},
  year={2024}
}

@inproceedings{zhou2024plankton,
  title={Plankton: Reconciling binary code and debug information},
  author={Zhou, Anshunkang and Ye, Chengfeng and Huang, Heqing and Cai, Yuandao and Zhang, Charles},
  booktitle={Proceedings of the 29th ACM International Conference on Architectural Support for Programming Languages and Operating Systems, Volume 2},
  pages={912--928},
  year={2024}
}

@book{cifuentes1994reverse,
  title={Reverse compilation techniques},
  author={Cifuentes, Cristina},
  year={1994},
  publisher={Queensland University of Technology, Brisbane}
}

@inproceedings{song2008bitblaze,
  title={BitBlaze: A new approach to computer security via binary analysis},
  author={Song, Dawn and Brumley, David and Yin, Heng and Caballero, Juan and Jager, Ivan and Kang, Min Gyung and Liang, Zhenkai and Newsome, James and Poosankam, Pongsin and Saxena, Prateek},
  booktitle={International conference on information systems security},
  pages={1--25},
  year={2008},
  organization={Springer}
}

@article{wong2025decllm,
  title={DecLLM: LLM-Augmented Recompilable Decompilation for Enabling Programmatic Use of Decompiled Code},
  author={Wong, Wai Kin and Wu, Daoyuan and Wang, Huaijin and Li, Zongjie and Liu, Zhibo and Wang, Shuai and Tang, Qiyi and Nie, Sen and Wu, Shi},
  journal={Proceedings of the ACM on Software Engineering},
  volume={2},
  number={ISSTA},
  pages={1841--1864},
  year={2025},
  publisher={ACM New York, NY, USA}
}

@inproceedings{he2018debin,
  title={Debin: Predicting debug information in stripped binaries},
  author={He, Jingxuan and Ivanov, Pesho and Tsankov, Petar and Raychev, Veselin and Vechev, Martin},
  booktitle={Proceedings of the 2018 ACM SIGSAC Conference on Computer and Communications Security},
  pages={1667--1680},
  year={2018}
}

@inproceedings{liu2025function,
  title={Function Renaming in Reverse Engineering of Embedded Device Firmware with ChatGPT},
  author={Liu, Puzhuo and Di, Peng and Jiang, Yu},
  booktitle={Proceedings of the 1st ACM SIGPLAN International Workshop on Language Models and Programming Languages},
  pages={57--65},
  year={2025}
}

@inproceedings{katz2018using,
  title={Using recurrent neural networks for decompilation},
  author={Katz, Deborah S and Ruchti, Jason and Schulte, Eric},
  booktitle={2018 IEEE 25th international conference on software analysis, evolution and reengineering (SANER)},
  pages={346--356},
  year={2018},
  organization={IEEE}
}

@inproceedings{schwartz2011q,
  title={Q: Exploit hardening made easy},
  author={Schwartz, Edward J and Avgerinos, Thanassis and Brumley, David},
  booktitle={20th USENIX Security Symposium (USENIX Security 11)},
  year={2011}
}

@inproceedings{meng2016binary,
  title={Binary code is not easy},
  author={Meng, Xiaozhu and Miller, Barton P},
  booktitle={Proceedings of the 25th International Symposium on Software Testing and Analysis},
  pages={24--35},
  year={2016}
}

@inproceedings{liu2026bit,
  title={BIT: Empowering Binary Analysis through the LLVM Toolchain},
  author={Liu, Puzhuo and Di, Peng and Xue, Jingling and Jiang, Yu},
  booktitle={Proceedings of the 24th ACM/IEEE International Symposium on Code Generation and Optimization},
  year={2026}
}

@inproceedings{mcintosh2011empirical,
  title={An empirical study of build maintenance effort},
  author={McIntosh, Shane and Adams, Bram and Nguyen, Thanh HD and Kamei, Yasutaka and Hassan, Ahmed E},
  booktitle={Proceedings of the 33rd international conference on software engineering},
  pages={141--150},
  year={2011}
}

@article{aidasso2025build,
  title={Build optimization: A systematic literature review},
  author={A{\"\i}dasso, Henri and Sayagh, Mohammed and Bordeleau, Francis},
  journal={ACM Computing Surveys},
  volume={58},
  number={1},
  pages={1--38},
  year={2025},
  publisher={ACM New York, NY}
}

@inproceedings{zhong2025understanding,
  title={Understanding compiler bugs in real development},
  author={Zhong, Hao},
  booktitle={2025 IEEE/ACM 47th International Conference on Software Engineering (ICSE)},
  pages={605--605},
  year={2025},
  organization={IEEE Computer Society}
}

@inproceedings{xu2023silent,
  title={Silent bugs matter: A study of $\{$Compiler-Introduced$\}$ security bugs},
  author={Xu, Jianhao and Lu, Kangjie and Du, Zhengjie and Ding, Zhu and Li, Linke and Wu, Qiushi and Payer, Mathias and Mao, Bing},
  booktitle={32nd USENIX Security Symposium (USENIX Security 23)},
  pages={3655--3672},
  year={2023}
}

@inproceedings{sun2016toward,
  title={Toward understanding compiler bugs in GCC and LLVM},
  author={Sun, Chengnian and Le, Vu and Zhang, Qirun and Su, Zhendong},
  booktitle={Proceedings of the 25th international symposium on software testing and analysis},
  pages={294--305},
  year={2016}
}

@inproceedings{eschweiler2016discovre,
  title={Discovre: Efficient cross-architecture identification of bugs in binary code.},
  author={Eschweiler, Sebastian and Yakdan, Khaled and Gerhards-Padilla, Elmar and others},
  booktitle={Ndss},
  volume={52},
  pages={58--79},
  year={2016}
}

@inproceedings{richter2023train,
  title={How to train your neural bug detector: Artificial vs real bugs},
  author={Richter, Cedric and Wehrheim, Heike},
  booktitle={2023 38th IEEE/ACM International Conference on Automated Software Engineering (ASE)},
  pages={1036--1048},
  year={2023},
  organization={IEEE}
}

@article{cifuentes1995decompilation,
  title={Decompilation of binary programs},
  author={Cifuentes, Cristina and Gough, K John},
  journal={Software: Practice and Experience},
  volume={25},
  number={7},
  pages={811--829},
  year={1995},
  publisher={Wiley Online Library}
}

@inproceedings{xie2024resym,
  title={Resym: Harnessing llms to recover variable and data structure symbols from stripped binaries},
  author={Xie, Danning and Zhang, Zhuo and Jiang, Nan and Xu, Xiangzhe and Tan, Lin and Zhang, Xiangyu},
  booktitle={Proceedings of the 2024 on ACM SIGSAC Conference on Computer and Communications Security},
  pages={4554--4568},
  year={2024}
}

@inproceedings{tan2024llm4decompile,
  title={LLM4Decompile: Decompiling Binary Code with Large Language Models},
  author={Tan, Hanzhuo and Luo, Qi and Li, Jing and Zhang, Yuqun},
  booktitle={Proceedings of the 2024 Conference on Empirical Methods in Natural Language Processing},
  pages={3473--3487},
  year={2024}
}

@misc{ghidra,
title = "Ghidra",
author= "NSA",
howpublished = "https://ghidra-sre.org",
year="2026",
note = {Accessed 2026-1-1}
}

@misc{guan2025benchmarkstillusefuldynamic,
      title={Is Your Benchmark (Still) Useful? Dynamic Benchmarking for Code Language Models}, 
      author={Batu Guan and Xiao Wu and Yuanyuan Yuan and Shaohua Li},
      year={2025},
      eprint={2503.06643},
      archivePrefix={arXiv},
      primaryClass={cs.SE},
      url={https://arxiv.org/abs/2503.06643}, 
}

@misc{adrp,
title = "ARM ADR/ADRP demos",
author= "Cliff Fan",
howpublished = "https://duetorun.com/blog/20230609/arm-adr-demo",
year="2023",
note = {Accessed 2026-1-1}
}

@misc{ida,
title = "IDA Pro",
author= "hex-rays",
howpublished = "https://hex-rays.com/ida-pro",
year="2026",
note = {Accessed 2025-1-1}
}

@Misc{retdec,
  title =        {RetDec},
  author =       {Avast},
  year =         2026,
  howpublished = {\url{https://github.com/avast/retdec/}},
  note = {Accessed 2026-1-1}
}

@inproceedings{shoshitaishvili2016state,
  title={{SoK: (State of) The Art of War: Offensive Techniques in Binary Analysis}},
  author={Shoshitaishvili, Yan and Wang, Ruoyu and Salls, Christopher and
          Stephens, Nick and Polino, Mario and Dutcher, Audrey and Grosen, John and
          Feng, Siji and Hauser, Christophe and Kruegel, Christopher and Vigna, Giovanni},
  booktitle={IEEE Symposium on Security and Privacy},
  year={2016}
}

@misc{binaryai,
  title={{BinaryAI}: Binary Code Analysis with Artificial Intelligence},
  author={{Tencent Security}},
  howpublished={\url{https://www.binaryai.net/}},
  year={2024}
}

@misc{frida,
  title={{Frida}: A World-Class Dynamic Instrumentation Toolkit},
  author={Ravnås, Ole André},
  howpublished={\url{https://frida.re/}},
  year={2024}
}

@misc{lima,
  title={{Lima}: Linux Virtual Machines},
  author={Akihiro Suda and {Lima contributors}},
  howpublished={\url{https://lima-vm.io/}},
  year={2024}
}

\end{document}